\documentclass[12pt]{article}
\usepackage{graphicx}
\usepackage{enumerate}
\usepackage{natbib}
\usepackage{color}
\usepackage{comment}
\usepackage{amsmath,amssymb,amsfonts}
\usepackage{url} 
\usepackage{setspace}

\usepackage{natbib}
\usepackage{url} 
\newcommand{\blind}{1}

\begin{document}

\def\spacingset#1{\renewcommand{\baselinestretch}%
{#1}\small\normalsize} \spacingset{1}


\if1\blind
{
  \title{\bf Tensor time series change-point detection in cryptocurrency network data}
  \author{Andreas Anastasiou
  \hspace{.2cm}\\
    Department of Mathematics and Statistics, University of Cyprus\\
    and \\
    Ivor Cribben\thanks{The second author was supported by the Natural Sciences and Engineering Research Council (Canada) grant RGPIN-2024-06102 and the Xerox Faculty Fellowship, Alberta School of Business.
} \\
    Department of Accounting and Business Analytics, Alberta School of Business,\\ University of Alberta}
  \maketitle
} \fi

\if0\blind
{
  \bigskip
  \bigskip
  \bigskip
  \begin{center}
    {\LARGE\bf Tensor time series change-point detection in cryptocurrency network data}
\end{center}
  \medskip
} \fi

\bigskip
\begin{abstract}
Financial fraud has been growing exponentially in recent years.  The rise of cryptocurrencies as an investment asset has simultaneously seen a parallel growth in cryptocurrency scams. To detect possible cryptocurrency fraud, and in particular market manipulation, previous research focused on the detection of changes in the network of trades; however, market manipulators are now trading across multiple cryptocurrency platforms, making their detection more difficult.  Hence, it is important to consider the identification of changes across several trading networks or a `network of networks' over time.  To this end, in this article, we propose a new change-point detection method in the network structure of tensor-variate data. This new method, labeled TenSeg, first employs a tensor decomposition, and second detects multiple change-points in the second-order (cross-covariance or network) structure of the decomposed data.  It allows for change-point detection in the presence of frequent changes of possibly small magnitudes and is computationally fast. We apply our method to several simulated datasets and to a cryptocurrency dataset, which consists of network tensor-variate data from the Ethereum blockchain. We demonstrate that our approach substantially outperforms other state-of-the-art change-point techniques, and the detected change-points in the Ethereum data set coincide with changes across several trading networks or a `network of networks' over time. Finally, all the relevant \textsf{R} code implementing the method in the article are available on
\url{https://github.com/Anastasiou-Andreas/TenSeg}.
\end{abstract}

\noindent%
{\it Keywords:} Matrix time series; tensor segmentation; isolate-detect; canonical polyadic decomposition; Ethereum blockchain data; Enron email dataset.

\vfill

\newpage
\spacingset{1.75} 
\section{Introduction}
\label{sec:intro}

There exists an extensive literature and a long history of research on change-points, beginning with the work of \cite{page}.  Change-point detection in one-dimensional data is the simplest and most well-studied scenario, where the goal is to identify abrupt shifts in a univariate time series or sequence. Classical methods, such as the CUSUM (cumulative sum) test, likelihood ratio tests, and Bayesian approaches, are widely used to locate changes in mean, variance, or distribution \citep{inclan1994,chen1997}. These techniques rely on statistical hypothesis testing or optimization frameworks to partition the data into homogeneous segments. One-dimensional change-point detection serves as the foundation for more complex problems, offering interpretable results and efficient computational solutions, making it applicable in fields such as quality control, econometrics, and environmental monitoring.  

When moving to multivariate data, the complexity increases as dependencies between variables must be taken into account. Methods such as Hotelling’s $T^{2}$ statistic, multivariate CUSUM, and subspace-based approaches are employed to detect changes in the joint distribution or covariance structure of the data \citep{galeano2007,cribben2012,cribben2013,gibberd2014,cho2015,kirch2015,cribben2017,barnett2016}. Unlike univariate cases, where changes are often scalar (e.g., mean shift), multivariate settings require detecting shifts in vectors or matrices, increasing computational and statistical challenges. Applications include fault detection in industrial systems, brain signal analysis, and financial market monitoring, where multiple correlated signals must be analyzed simultaneously.  

In high-dimensional settings, where the number of features ($p$) may exceed the number of observations ($n$), traditional change-point methods often fail due to the curse of dimensionality. Sparse regularization, dimensionality reduction, and graph-based techniques become essential to identify meaningful changes without overfitting. Recent advances leverage sparsity assumptions, such as detecting changes in only a subset of dimensions, or employ machine learning techniques such as deep learning for feature extraction.  Addressing these challenges requires a blend of statistical rigor and scalable algorithms to ensure robustness in large-scale applications.
The latest work on the topic includes methods for change-points in graph summary statistics \citep{ofori2021}, change-points in the clustering structure for multivariate high-dimensional time series \citep{ondrus2021}, multiple change-points in the second-order structure of multivariate, possibly high-dimensional time series \citep{anastasiou2022}, and change-points in the covariance structure of moderate-dimensional time series \citep{killick2023}. \cite{zhang2024} used a tensor-based embedding mode for changes in the network structure in a unilayer network.  Finally, \cite{huang2022,qin2023} studied mean changes in tensor variate data.

Financial fraud is a deceptive practice employed with the intention of illicitly obtaining money or sensitive information from unsuspecting individuals or organizations. These fraudulent acts (or scams) have become increasingly prevalent in the digital age, as technology advancements and the widespread use of the Internet have provided perpetrators with new avenues to exploit easily. Financial scams cover a wide range of activities, including pyramid schemes, Ponzi schemes, identity theft, phishing, and fake investment opportunities.  The perpetrators typically employ techniques to build trust and create an illusion of legitimacy and credibility, luring their victims into parting with money or divulging sensitive information. Financial scams not only cause significant financial losses to individuals and organizations, but also erode trust in financial systems, undermining economic stability. Hence, it is crucial for individuals, businesses, and regulatory authorities to remain vigilant and implement robust preventive measures to combat this pervasive issue.

Cryptocurrencies are a relatively new form of digital or virtual currency. Recently, cryptocurrencies have gained significant attention and popularity, promising increased security, privacy, and potential for financial inclusion.  Cryptocurrencies, such as Bitcoin \citep{nakamoto} and Ethereum \citep{buterin}, are created through a process known as mining, where complex mathematical problems are solved to validate and record transactions on the blockchain. The decentralized nature of cryptocurrencies means that they are not controlled by any central authority, such as a government or financial institution, which has both positive and negative implications. A positive implication is that the decentralized structure allows greater transparency and eliminates the need for intermediaries in financial transactions. A negative implication is that it raises concerns about regulatory oversight, market volatility, and the potential for illicit activities, including money laundering and financing of illegal operations. Despite these challenges, cryptocurrencies have received an incredible amount of investment and have generated significant interest as an alternative currency and a potential store of value. Ongoing research and regulatory efforts are necessary to understand and address the opportunities and risks associated with cryptocurrencies, as they continue to shape digital finance \citep{feinstein}.

Cryptocurrency scams have emerged as a major concern within the digital finance landscape.  Chain Analysis reported that USD\$24.2 billion worth of cryptocurrency crime occurred in 2023.  These scams often employ sophisticated tactics to deceive investors and exploit their lack of understanding or knowledge about cryptocurrencies. For example, scammers may create fraudulent websites or social media accounts that mimic legitimate cryptocurrency platforms, luring unsuspecting victims to invest in nonexistent or worthless digital assets. In addition, pump-and-dump schemes, in which scammers artificially inflate the price of a particular cryptocurrency before selling their holdings at a profit, have become increasingly prevalent.  \cite{trozze} concluded that the academic literature has identified 29 different types of cryptocurrency fraud and ranked pump-and-dump schemes and ransomware as the most profitable and feasible threats.  The lack of regulatory oversight and the pseudonymous nature of cryptocurrency transactions make it challenging to identify and hold perpetrators accountable.  

The literature on cryptocurrency fraud detection encompasses a diverse range of methodologies and approaches aimed at identifying and mitigating fraudulent activities within blockchain-based systems. \cite{conti} provide a survey of techniques for detecting fraudulent transactions, emphasizing the importance of machine learning algorithms in enhancing detection accuracy. Several studies, such as \cite{monamo}, focus on unsupervised learning methods, leveraging clustering and anomaly detection to identify suspicious patterns in transaction data. Furthermore, \cite{alrubaian} propose a reputation-based trust management system that uses user behavior analysis to prevent fraud in cryptocurrency exchanges. The integration of graph-based techniques, as discussed by \cite{weber}, highlights the effectiveness of analyzing transaction networks to uncover illicit activities or detecting market manipulation. Despite the advancements, ongoing research continues to address the challenges posed by the evolving nature of fraudulent behaviors in cryptocurrencies.

Statistical and machine learning methods play a crucial role in detecting fraud in cryptocurrency transactions, providing valuable tools to identify anomalous patterns and suspicious activities within the vast and complex blockchain networks \citep{weber}. One common approach involves the analysis of transactional data using advanced statistical models and algorithms. Changes in transaction frequency, size, or geographical location can be flagged for further investigation \citep{wu,bartoletti}.  Furthermore, clustering algorithms can group transactions that exhibit similar characteristics, helping to uncover hidden connections among potentially fraudulent activities \citep{phillips}.  Networks are a general language for describing interacting systems in the real world, and a considerable part of the existing work on cryptocurrency transactions is studied from a network perspective.  \cite{wu} present the background information on the analysis of cryptocurrency transaction networks and review existing research in terms of three aspects: network modeling, network profiling, and network-based detection. For each aspect, they introduce the research issues, summarize the methods, and discuss the results and findings in the literature.  \cite{weber} show that the transactions in coins on the Bitcoin graph provide important insights into money laundering schemes. As criminal, fraudulent and illicit activities such as market manipulation on blockchains continue to increase, cryptocurrency criminals increasingly employ cross-cryptocurrency trades to hide their identity \citep{nelson}. \cite{yousaf} studied whether money can be traced as it moves not only within the ledger of a single cryptocurrency but if it can in fact be traced as it moves across ledgers.
However, while there exist some statistical methods for change-points in single cryptocurrency networks \citep{tan,james,assaf}, there is not, to the best of our knowledge, any change-point method to detect change-points (or market manipulation) in several cryptocurrency networks (or ledgers).

To this end, we introduce a new method, called TenSeg, that detects change-points across multiple time-evolving blockchain transaction networks (or tokens). The nature of change-point detection in TenSeg is offline, in the sense that all the data are available prior to segmentation taking place. However, the pseudo-sequential nature of the detection process in TenSeg (see Section \ref{sec:ccid} for more details) makes the algorithm easily adaptable to an online framework that could potentially be used by organizations to identify (possible) market manipulation promptly after it occurs.  We apply TenSeg to an Ethereum data set, which includes only tokens that have more than \$100M in market value, as reported by the EtherScan.io online explorer. The processed data consist of 6 tokens, namely ``bat'', ``golem'', ``mcap'', ``statusnetwork'', ``storj'', and ``tenxpay'' over a period of 300 days from July 11, 2017 to May 7, 2018. Our objective is to study whether the investment or trading activity pattern changes between the 100 investors across these token networks.  The TenSeg method assumes a tensor variate structure of the data, as crime on blockchain transaction networks, such as market manipulation or money laundering, usually involves multiple parties who have the opportunity to move funds across multiple cryptocurrency tokens. TenSeg first employs a tensor decomposition, and then detects multiple change-points in the second-order (cross-covariance or network) structure of the decomposed data.  As described above these change-points could be due to various factors including a pump-and-dump scheme \citep{xu2019}.

Our contributions are as follows. First, from a statistical application perspective, this is only work thus far that has explored change-points across multiple cryptocurrency (Ethereum) token networks. Second, from a statistical methodology perspective, to the best of our knowledge, TenSeg is the first article on change-point detection in the network structure of tensor-variate data (\citealp{huang2022} and \citealp{qin2023} only considered mean changes in tensor-variate data).  Third,  in the absence of state-of-the-art change-point methods for tensor data, we benchmark TenSeg against other recent change-point methods in the second-order structure after applying the tensor decomposition.  Fourth, while TenSeg uses a canonical polyadic decomposition, we show that our results are robust to this choice.  Fifth, the pseudo-sequential nature of the isolation part in our algorithm (see Section \ref{sec:ccid} for more details) makes the proposed methodology easily adaptable to an online change-point framework, where the aim is to detect changes while new observations are arriving. Such an extension will serve as a powerful tool for identifying, for example, potential cryptocurrency scams (such as market manipulation) promptly as data arrives. Sixth, we show that TenSeg can work well in the presence of autocorrelation, a common impediment for many change-point detection methods. Seventh, we show that TenSeg is generalizable by applying it to the Carnegie Mellon University CALO Project Enron email dataset, and we detail new perspectives on the data set.

The remainder of this article is organized as follows.  We introduce and motivate our Ethereum blockchain data in Section~\ref{sec:data}.  We describe our new method, TenSeg, in Section~\ref{sec:methods}. The simulation study, along with the results obtained, can be found in Section~\ref{sec:sims}. The performance of the proposed method on real data is detailed in Section~\ref{sec:real_data_results}. A discussion of the robustness of the method, as well as some limitations, is described in Section~\ref{sec:discussion}.  Finally, we conclude in Section~\ref{sec:conc}.
\section{Ethereum blockchain data}
\label{sec:data}
Ethereum is a technology that is home to digital money, global payments, and applications. The Ethereum community has built a booming digital economy with bold new ways for creators to earn money online. It is open to everyone, wherever you are in the world, assuming you have access to the Internet.  The Ethereum blockchain was created in 2015 to implement smart contracts, which are Turing-complete software codes that execute user-defined tasks.  Ethereum smart contracts serve as the backbone of token networks, providing the infrastructure for the seamless operation of decentralized applications. These contracts enable the creation, deployment, and execution of digital agreements that control the behavior of tokens within the network. Using Ethereum's blockchain technology, token networks ensure transparency, security, and immutability in transactions, fostering trust among participants.  Smart contracts within these networks facilitate the automated execution of predefined conditions, enabling the seamless transfer and management of tokens without the need for intermediaries. As a result, Ethereum smart contracts play a pivotal role in the efficient functioning and expansion of tokenized ecosystems.

A token network is a directed weighted graph with an edge denoting the transferred token value.  Token networks are particularly valuable because each token naturally represents a network layer with the same nodes (addresses of investors) appearing in the networks (layers) of multiple tokens. Our Ethereum data set used in this article consists of extracted traded token networks between investors from the publicly available Ethereum blockchain. We use the normalized number of transactions between nodes as edge weights. Although address creation is cheap and easy, most blockchain users have the same address for a long period and use it to trade multiple tokens. As a result, the address appears in networks of all the tokens it has traded. 

From our Ethereum data set, we only include tokens reported by the EtherScan.io online explorer to have more than \$100M in market value.  The tokens have on average a history of 271 days (minimum and maximum of 137 and 443 days, respectively).  Each token has a different creation date, hence token networks have non-identical lifetime intervals. Hence, in our data preprocessing, we match the tokens to maximize the overlap of trading times. The processed data consist of 6 tokens, namely ``bat'', ``golem'', ``mcap'', ``statusnetwork'', ``storj'', and ``tenxpay'' over a period of 300 days from July 11, 2017 to May 7, 2018. We observe trading activity between 100 investors, which means that the tensor object obtained for this data set is a tensor of dimensionality $6 \times 100 \times 100 \times 300$ (\# tokens $\times$ investor network $\times~T$). Our objective is to study whether the investing activity (weighted networks in a tensor object) pattern changes between the 100 investors for the aforementioned 6 tokens during this 300-day period. As we focus on trading activity between investors and not on Ethereum prices, the inherent volatility of cryptocurrency prices is not an issue. 

\section{Methods} \label{sec:methods}

\subsection{Tensors}
A tensor is a multi-dimensional array, with the modes of a tensor corresponding to the dimensions of the array.  For example, a vector is a $1$-tensor, a matrix a $2$-tensor, and tensors with 3 or more modes are called higher-order tensors.  Suppose that $K$ and $n_{1}\times n_{2}\times ... \times n_{K}$ denote the number of modes of a tensor and the extents of the modes associated with the $K$-tensor, respectively.  The unfolding or flattening of a $K\ge 3$ tensor corresponds to representing the tensor as a matrix or a vector.  One can select exactly which modes of the general $K$-tensor to map onto the rows and columns to represent the matrix.  One common unfolding is called the $k$-mode matricization/unfolding.  For a tensor $\mathcal{X} \in \mathbb{R}^{n_{1}\times n_{2}\times...\times n_{K}}$, the unfolding in the $k$th mode can be denoted as 
$\mathcal{X}_{(k)} \in \mathbb{R}^{n_{k}\prod_{j\neq k} n_{j}}$.
For example, for a 3-tensor, there are three $k$-mode unfoldings, denoted $\mathcal{X}_{(1)},\mathcal{X}_{(2)}, \mathcal{X}_{(3)}$.  The $k$-mode product is the multiplication of a $K$-tensor $\mathcal{X} \in \mathbb{R}^{n_{1}\times n_{2}\times...\times n_{K}}$ and a matrix $M \in \mathbb{R}^{J\times n_{k}}$, which results in a $K$-tensor in $\mathbb{R}^{n_{1}\times n_{2}\times n_{k-1}\times J\times n_{k+1} ...\times n_{K}}$.  Elementwise, this is defined as \linebreak $(\mathcal{X}\times_{k} M)_{i_{1},...,i_{k-1},j,i_{k+1},...,i_{K}} = \sum_{i_k = 1}^{n_{k}}\mathcal{X}_{i_{1},...,i_{k}} . M_{j,i_{k}}.$ Evidently, the $k$-mode product is closely related to the $k$-mode unfolding.  In particular, $\mathcal{Y} = \mathcal{X} \times_{k} M \Leftrightarrow \mathcal{Y}_{(k)} = M.\mathcal{X}_{(k)},$ where $.$ denotes the usual matrix multiplication.

In this article, we have the following hypotheses for the detection of change-points in the network structure $\Theta$ of a tensor $\mathcal{X}$:
\begin{equation*}
\begin{aligned}
H_0 &: \Theta_t = \Theta_{t+1}, \quad \forall t \in \{1, \dots, T-1\}, \\
H_A &: \exists \, r_1, r_2, \dots, r_N \in \{1, \dots, T-1\} \text{ such that } 
\Theta_{r_i} \neq \Theta_{r_i+1}, \quad \forall i \in \{1, \dots, N\}.
\end{aligned}
\end{equation*}

\subsection{Tensor decomposition}
A decomposition of a higher-order tensor is often called multi-way analysis or multi-linear models.  In the proposed TenSeg methodology, we use the Canonical Polyadic (CP) decomposition, which was introduced in \cite{Hitchcock}. A $K$-tensor $\mathcal{X} \in  \mathbb{R}^{n_{1}\times n_{2}\times ...\times n_{K}}$ is denoted rank-1 if it can be expressed as an outer product of $K$ vectors. Alternatively, the rank of a $K$-tensor is defined as the minimum value, $r$, such that the $K$-tensor can be expressed as the sum of rank-1 tensors
\begin{equation*}
\mathcal{X} = \sum_{l=1}^{r}v_{1l}\circ v_{2l}\circ ...\circ v_{Kl}, ~\text{where}~ v_{kl}\in \mathbb{R}^{n_{k}}, 1\le l \le r, 1\le k\le K.
\end{equation*}
\cite{kolda2003} showed that the existence and form of an optimal lower-rank approximation does not generalize to $K$-tensors for $K\ge 3$.
The CP decomposition approximates $\mathcal{X}$ with a rank-$r$ tensor $\hat{\mathcal{X}}$, where $r$ is specified a priori. The objective is to construct a rank-$r$ tensor that minimizes the Frobenius
norm ($\underset{{~\text{all}~v_{kl}}}{\min}||\mathcal{X} - \hat{\mathcal{X}}||_{F}$) of the difference between $\hat{\mathcal{X}}$ and $\mathcal{X}$, with
\begin{align} \label{eq:tensor_decomp}
\mathcal{X} = \sum_{l=1}^{r}v_{1l}\circ v_{2l}\circ ...\circ v_{Kl} 
            &= \sum_{l=1}^{r} \lambda_{l} u_{1l}\circ u_{2l}\circ ...\circ u_{Kl},\;\; u_{Kl} = \frac{v_{kl}}{||v_{kl}||}\\
            &= \Lambda \times_{1} U_{1} \times_{2} U_{2} \times_{3} ... \times_{K} U_{K}, 
\end{align}
\noindent where $\times_i$ represents the mode-$i$ tensor multiplication, $U_{k} = [u_{k1} u_{k2} ... u_{kr}] \in \mathbb{R}^{n_{k}\times r}$, and $\Lambda \in \mathbb{R}^{r\times r\times r}$ is a $3$-tensor that contains the $\lambda$'s on the super-diagonal and $0$ elsewhere. The equivalence in equation (\ref{eq:tensor_decomp}) arises as the norm information is stored in the $\lambda_{r}$'s.  In addition, we can store the $u_{Kl}$ vectors as factor matrices $U_{k}$, for each $k=1,...K$, which are not orthogonal. 

Assuming the number of components is fixed, there are many  algorithms available to compute the CP decomposition. We use the \textit{cp} function from the \textbf{rtensor} {\textsf{R}} package \citep{li2018}, which implements the ``workhorse'' algorithm of CP, the classical alternating least squares (ALS) method proposed by \cite{carroll}.  The ALS method is simple, but it can take many iterations to converge.  It is also not guaranteed to converge to a global minimum, only to converge to a solution where the objective function ceases to decrease.  The final estimate depends on the initialization.  For more details on the ALS method and its many extensions, see \cite{kolda2009}.
\subsection{Methods for choosing the optimal rank} \label{sec:normo}
\begin{figure}[!ht]
\centering
\includegraphics[height=6cm, width=13cm]{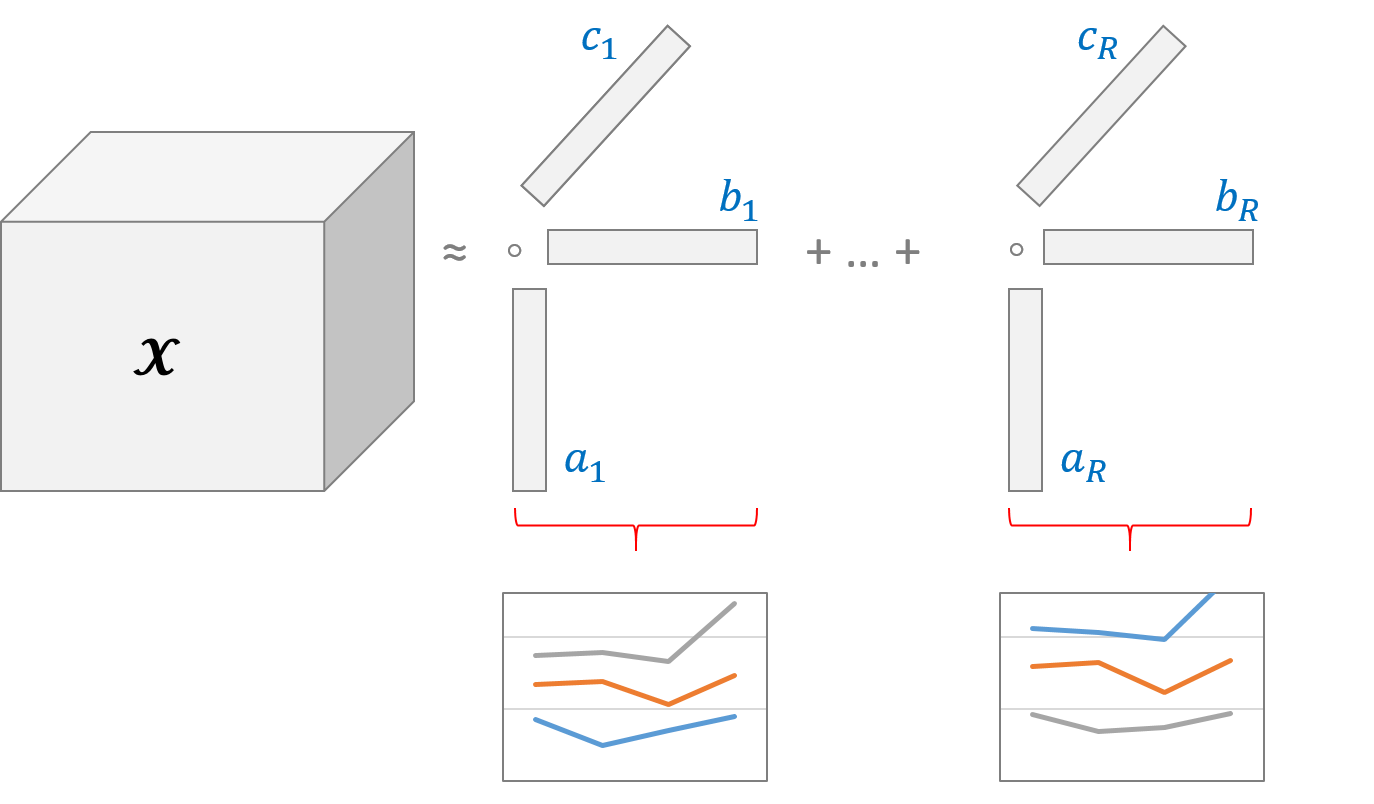}
\caption{An illustration of the output from a Canonical Polyadic (CP) decomposition of a tensor $\mathcal{X}$ into components.}
\label{fig:CP}
\end{figure}

A CP decomposition generates a set of components (also called factors or concepts) that model the patterns in the data. One of the properties of CP is that it allows for the discovery of overlapping components.  For example, for time-varying network data, CP may find patterns that share active nodes, hence it may provide redundant information.  To address the problem of redundancy and choosing the optimal rank in a CP decomposition, several methods have been proposed, including CORCONDIA \citep{bro2003}, DIFFerence in Fit \citep{timmerman2000}, Numerical Convex Hull method \citep{ceulemans2006}, and the Automatic Relevance Determination \citep{morup2009}.  In this work, we use the Redundant Model Order estimator (NORMO: \citealp{fernandes2020}) method to select the optimal rank in a CP decomposition due to its
superior performance. The setup is as follows.  Suppose that we have a tensor of three dimensions $\mathcal{X} \in \mathbb{R}^{N_{1}\times N_{2}\times N_{3}}$, and after applying a CP decomposition, the objective is to find $R$ vectors $\mathbf{a}_{r} \in \mathbb{R}^{N_{1}}$, $\mathbf{b}_{r} \in \mathbb{R}^{N_{2}}$, $\mathbf{c}_{r} \in \mathbb{R}^{N_{3}}$ such that
$\mathcal{X}(n_{1},n_{2},n_{3}) = \sum_{r=1}^{R} \mathbf{a}_{r}(n_{1})\mathbf{b}_{r}(n_{2})\mathbf{c}_{r}(n_{3})$, where the vectors can be grouped into factor matrices $A \in \mathbb{R}^{N_{1}\times R}$, $B \in \mathbb{R}^{N_{2}\times R}$, and $C \in \mathbb{R}^{N_{3}\times R}$.  The CP decomposition can be written as $\mathcal{X} = \sum_{r=1}^{R} \mathbf{a}_{r}\circ \mathbf{b}_{r}\circ \mathbf{c}_{r}$, where $\circ$ denotes the vector of outer product. The number $R$ is called the rank of the CP model. In Figure \ref{fig:CP}, we illustrate the output of a CP decomposition. Each component $r$ is defined by the triple \{$\mathbf{a}_{r},\mathbf{b}_{r}, \mathbf{c}_{r}$\} and maps a pattern of the tensor data. For example, in a person$\times$person$\times$time tensor data set modelling a time-evolving network, each component maps a communication pattern.

Given two scores defined by \{$\mathbf{a}_{{r}_{1}},\mathbf{b}_{{r}_{1}},\mathbf{c}_{{r}_{1}}$\} and \{$\mathbf{a}_{{r}_{2}},\mathbf{b}_{{r}_{2}},\mathbf{c}_{{r}_{2}}$\}, NORMO assigns a similarity metric to the scores where high similarity indicates redundancy.  NORMO is the average correlation across modes ($c_{A}$), in which we
define $c_{1}(r_{1},r_{2}) = |\text{cor}(\mathbf{a}_{{r}_{1}},\mathbf{a}_{{r}_{2}})|$; $c_{2}(r_{1},r_{2}) = |\text{cor}(\mathbf{b}_{{r}_{1}},\mathbf{b}_{{r}_{2}})|$; $c_{3}(r_{1},r_{2}) = |\text{cor}(\mathbf{c}_{{r}_{1}},\mathbf{c}_{{r}_{2}})|$, where cor($\mathbf{u}, \mathbf{v}$) denotes the correlation between the vectors
$\mathbf{u}$ and $\mathbf{v}$. Then, the correlation between the components $r_{1}$ and $r_{2}$ is defined as the average correlation $c_{A}(r_{1},r_{2}) = \frac{c_{1}(r_{1},r_{2})+c_{2}(r_{1},r_{2})+c_{3}(r_{1},r_{2})}{3}$.
The components $r_{2}$ and $r_{3}$ are said to be redundant if $c_{A}(r_{1},r_{2}) > \delta$. The inputs in the algorithm are the maximum number of
components to be tested and the correlation threshold to be
used is $\delta > 0$, with the default value in the NORMO algorithm set at 0.7. We use NORMO to select the number of components in the Ethereum blockchain data, while in the simulation study we show that our method is robust to misspecification to the number of components selected. 

\subsection{Cross-covariance isolate detect} \label{sec:ccid}
Once the decomposition of the tensor has been performed on the tensor data, $\mathcal{X}$, we apply the cross-covariance isolate detect (CCID: \citealp{anastasiou2022}) method to the decomposed matrix data set, $\mathbf{X}$. CCID is an algorithm for the detection of changes in the second-order (cross covariance or network) structure of a multivariate, possibly high-dimensional, time series. It starts by constructing appropriate wavelet-based local periodogram sequences from the $p$-variate locally stationary wavelet time series, $\mathbf{X}$, that is, $\mathbf{X}$ is transformed to periodograms and cross-periodograms creating a multiplicative model
\begin{equation}
\label{our_model}
Y_{t,T}^{(k)} = \sigma^{(k)}_{t,T}\left((Z_{t,T})^{(k)}\right)^2, \quad t=1,2,\ldots,T, \quad k=1,2,\ldots,d,
\end{equation}
where $T$ is the total length of the multivariate time series, which is of dimensionality $d$, and $Y_{t,T}^{(k)}$ are the observed data related to the $k^{\textsuperscript{th}}$ data sequence at time point $t$. Regarding $\sigma_{t,T}^{(k)}$, this is a piecewise-constant mean function and $Z_{t,T}^{(k)}$ is a sequence of (possibly) autocorrelated standard normal random variables. 
Changes in the network structure of the tensor, $\mathcal{X}$, correspond to changes in the second-order (or cross-covariance) structure of the data in $\mathbf{X}$, which in turn correspond to changes in the mean structure of $\sigma_{t,T}^{(k)}$. Therefore, our objective is to detect the number and the location of changes in the mean of the piecewise-constant $\sigma^{(k)}_{t,T}$, for any $k \in \left\lbrace 1,2,\ldots,d\right\rbrace$, with each change-point being possibly shared by more than one of the component data sequences in the matrix $\mathbf{X}$. Towards this, and with $\bar{Y}_{t1,t2}^{(k)} := \frac{1}{t_2 - t_1 + 1}\sum_{t = t_1}^{t_2}Y_{t,T}^{(k)}$, the scaled CUSUM statistic is defined as
\begin{equation}
\label{CUSUM}
\tilde{Y}_{s,e}^{b, (k)} = \left(\bar{Y}_{s,e}^{(k)}\right)^{-1}\sqrt{\frac{(e-b)(b-s+1)}{e-s+1}}\left|\bar{Y}_{s,b}^{(k)} - \bar{Y}_{b+1,e}^{(k)}\right|
\end{equation}
where $1\leq s \leq b < e\leq T$. The division by the sample mean of the observations $Y_{t,T}^{(k)}$, with $t \in [s,e]$, is necessary in multiplicative settings such as \eqref{our_model} to make the results independent of the magnitude of $\sigma_{t,T}^{(k)}$. More information on \eqref{CUSUM} can be found in \cite{inclan1994} and \cite{cho2015}.

Algorithmically, we rely on the change-point isolation principle as a first step; the relevant methodology, labeled Isolate-Detect (ID), was first developed in \cite{anastasiou2019} for the univariate offline change-point detection setting, while its multivariate extension, named MID, can be found in \cite{anastasiou2023}. 
The relevant design is that for $d$ observed data sequences each of length $T$, and with a suitably chosen positive integer $\lambda_T$, we first create two ordered sets of $K = \left\lceil T/\lambda_T \right\rceil$ right- and left-expanding intervals. More precisely, for
$i = 1, \ldots, K$, the $i^{\rm th}$ right expanding interval is $R_i = [1, \min\{i\lambda_T, T\}]$ while the $i^{\rm th}$ left-expanding interval is $L_i = [\max\{1, T - i\lambda_T + 1\}, T]$. At each step, we expand the intervals by $\lambda_T$ which is assumed to be smaller than the minimum distance between two consecutive change-points; this is where isolation stems from as well, which in order to be ensured in practice, $\lambda_T$ can be taken to be as small one or equal to 1. If $\lambda_T > 1$, then isolation is guaranteed with high probability. The intervals are collected in the ordered set $S_{RL} = \left\lbrace R_1, L_1, R_2, L_2, \ldots,R_K, L_K\right\rbrace$.

As a next step, for each point $b$ in the interval $R_1$ and for each $k = 1,2 \ldots, d$, we calculate the values of $\tilde{Y}_{1,\lambda_T}^{b,(k)}$ as in equation \eqref{CUSUM}.  We aggregate the information of each of the $d$ data sequences using a mean-dominant norm $\ell(\cdot)$; the mean-dominance property is such for any $d \in \mathbb{Z}^{+}$ and $\forall \boldsymbol{x} \in \mathbb{R}^{d}$ with $x_i \geq 0, i = 1,\ldots, d$, it holds that $\ell(\boldsymbol{x}) \geq \frac{1}{d}\sum_{i=1}^{d}x_i$. We focus on the $\ell_{2}$ and $\ell_\infty$ mean-dominant norms; more specifically, for $\ell_2$ we define ${\tilde{U}_{s,e}^{b,\ell_2}} = d^{-1/2}\sqrt{\sum_{k=1}^{d}\left(\tilde{Y}_{s,e}^{b,(k)}\right)^2}, \quad b \in [s,e)$ and we set $b^{*} := \underset{b \in R_1}{\rm argmax}~\tilde{U}_{1,\lambda_T}^{b,\ell_2}$. If this value exceeds a certain threshold, denoted by $\zeta_{T,d}^{\ell_2}$, then it is taken as a change-point. If not, then the process tests the next interval in $S_{RL}$, which in this case would be $L_1$. Upon detection, we make a new start from the end-point (or start-point, respectively) of the right- (or left-, respectively) expanding interval where the detection occurred.  At some point, we work only in intervals $[s,e]$ that do not contain any other change-points and the process stops. The $\ell_{\infty}$ approach works similarly; however, we now use $\tilde{U}_{s,e}^{b,\ell_\infty} = \max_{k \in \left\lbrace 1,2,\ldots, d\right\rbrace}\left\lbrace \left|\tilde{Y}_{s,e}^{b,(k)}\right|\right\rbrace, \quad b \in [s,e)$ and we set $\tilde{b} := \underset{b \in R_1}{\rm argmax}~\tilde{U}_{1,\lambda_T}^{b,\ell_{\infty}}$. If this value exceeds a threshold, denoted by $\zeta_{T,d}^{\ell_{\infty}}$, then it is taken as a change-point. We highlight that both $\zeta_{T,d}^{\ell_2}$ and $\zeta_{T,d}^{\ell_\infty}$ are of the form $C\sqrt{\log (Td^{1/4})}$, where $C$ is a positive constant; its choice in the case of structural changes in a multivariate data sequence is discussed in \cite{anastasiou2023}. 

A stopping rule, other than thresholding, is based on the optimization of a model selection criterion. In this setup, we begin by overestimating the number of change-points by choosing a suboptimal (lower) threshold value in the $\ell_2$ or the $\ell_{\infty}$ methods explained above. The estimated change-points are sorted in an increasing order in the set $\tilde{S} = \left\lbrace \tilde{r}_1, \tilde{r}_2, \ldots, \tilde{r}_{\tilde{N}}\right\rbrace$, for $\tilde{N} \geq N$ (the estimated number of change-points is greater than the true number of change-points). The next step is to run a change-point removal process using a joint approach for the $d$ data sequences.  For $\tilde{r}_0 = 0$ and $\tilde{r}_{\tilde{N}+1} = T$, we collect the triplets $\left(\tilde{r}_{j-1}, \tilde{r}_j, \tilde{r}_{j+1}\right)$ and calculate
$CS^*(\tilde{r}_j):= \max_{k=1,2,\ldots,d}\left\lbrace\tilde{Y}_{\tilde{r}_{j-1} + 1, \tilde{r}_{j+1}}^{\tilde{r}_j, (k)}\right\rbrace$. We then define $m = {\rm argmin}_{j}\left\lbrace CS^*(\tilde{r}_j) \right\rbrace$ and $\tilde{r}_{m}$ is labeled the least important detection and is removed from the set $\tilde{S}$. We relabel the remaining estimates (in increasing order) in $\tilde{S}$, and repeat this estimate removal process until $\tilde{S}$ becomes the empty set. After this, the vector $\boldsymbol{b} = \left(b_1,b_2,\ldots,b_{\tilde{N}}\right)$
represents the collection of estimates where $b_{\tilde{N}}$ is the estimate that was first removed from $\tilde{S}$, therefore the least important, $b_{\tilde{N}-1}$ is the estimate that was removed next, and so on. The vector $\boldsymbol{b}$ is called the solution path, and we define the collection of models $\left\lbrace\mathcal{M}_j\right\rbrace_{j = 0,1,\ldots,\tilde{N}}$ where $\mathcal{M}_{0} = \emptyset$ and $\mathcal{M}_j = \left\lbrace b_1,b_2,\ldots,b_j\right\rbrace$.  We choose the model that minimizes an appropriately chosen information criterion. For more details on the construction of the information criterion using a pseudo-likelihood in the case of a multivariate data sequence of a multiplicative structure, see Section 2.7 of \cite{anastasiou2022}.
\section{Simulations} \label{sec:sims}
\subsection{Description of the settings}\label{subsec:description}
The objective of the simulation study is to mimic the properties of our Ethereum blockchain data under various settings. We cover a wide range of different scenarios including several data structures, and varying the number of true change-points, the minimum distance between consecutive change-points, and the change magnitudes, to explore the flexibility of the proposed method.  For $T$ the time series length, the settings considered are:

\noindent CP0: $T = 200$ with no change-points. 

\noindent CP1: $T = 200$ with 1 change-point at time $t=100$.

\noindent CP4: $T = 300$ with 4 change-points at times $t= 100, 150, 200, 250$.

\noindent CP10: $T = 660$ with 10 change-points at times $t= 60, 120, \ldots, 600$.

To generate the $K$-tensor data, $\mathcal{X}$, we assume the following representation
\begin{equation} \label{eq:precision}
\mathcal{X} \times_{1} \mathbf{\Psi}_{1} + ... + \mathcal{X} \times_{K} \mathbf{\Psi}_{K} = \mathcal{T},
\end{equation}
\noindent where $\mathbf{\Psi}_{k} \in \mathbb{R}^{n_{k}\times n_{k}}, k =1,...,K$, are sparse precision matrices (symmetric and positive definite) and $\mathcal{T}$ is a random tensor of the same order as $\mathcal{X}$. Eq. \eqref{eq:precision} is called the Sylvester tensor equation \citep{wang2020}.  This equation arises in finite difference discretization of linear partial equations in high dimension.  We can show that \eqref{eq:precision} is equivalent to $\left(\bigoplus_{k = 1}^{K}  \mathbf{\Psi}_{k}\right) \text{vec}(\mathcal{X}) = \text{vec}(\mathcal{T})$, where \text{vec} indicates the vectorized version of the tensor. To generate the simulated tensor data, $\mathcal{X}$, we take $K = 4$ and $n_1 = n_2 = n_3 = 20$, while $n_4 = T$, where $T$ is the total time series length for each of the settings described above. For the tensor $\mathcal{T}$, we take $\text{vec}(\mathcal{T}) \sim \mathcal{N}(\boldsymbol{0}, \boldsymbol{I}_m)$, where $m = \prod_{k=1}^{K}n_k$. Lastly, we need to specify $\mathbf{\Psi}_{k}$.  For simplicity, we generate $\mathbf{\Psi}_{k}$ as identical $20 \times 20$ precision matrices based on three different structures: 1) AR1($\rho$), 2) Star-Block (SB), and 3) Erdos-Renyi (ER) random graph models. When the $\mathbf{\Psi}_{k}$'s are generated from an AR1($\rho$) random graph model, the covariance matrix is of the form $\mathbf{A} = (\rho^{|i-j|})_{ij}$. In the CP0 setting, we select $\rho=0.2$, in CP1 $\rho = 0.2$ for the first 100 time points, while $\rho = 0.8$ for the second 100 time points. For CP4 and CP10, the value of $\rho$ alternates between 0.2 and 0.8 in ABABA and ABABABABABA type settings, respectively. 

When the $\mathbf{\Psi}_{k}$'s are generated from a Star-Block (SB) random graph model, which is a block-diagonal covariance matrix, each block’s precision matrix corresponds to a star-structured graph with $(\mathbf{\Psi}_{k})_{ij}=1$. Then, for $\rho \in (0,1)$, we have $\mathbf{A}_{ij} =\rho$ if $(i,j) \in E$ and $\mathbf{A}_{ij} =\rho^{2} \notin E$, where $E$ is the corresponding edge set.  In the CP0 setting, we select $\rho=0.2$ and the number of subgraphs equal to $4$, in CP1 $\rho = 0.8$ and the number of subgraphs equal to $4$ for the first 100 time points, while $\rho = 0.2$ and the number of subgraphs equal to $2$ for the second 100 time points. For CP4 and CP10, the value of $\rho$ alternates between 0.8 and 0.2, and the number of subgraphs is from $4$ to $2$ in ABABA and ABABABABABA type settings, respectively. 

When the $\mathbf{\Psi}_{k}$'s are generated from an Erdos-Renyi random graph (ER), the precision matrix is initialized at $\mathbf{A} = 0.25\mathbf{I}$, and $d$ edges are randomly selected. For the selected edge $(i, j)$, and for $0 \le a_1 < a_2$ we randomly choose $\gamma \in [a_1, a_2]$ and update $\mathbf{A}_{ij} = \mathbf{A}_{ji} \to \mathbf{A}_{ij} - \gamma$ and $\mathbf{A}_{ii} \to \mathbf{A}_{ii} + \gamma$, $\mathbf{A}_{jj} \to \mathbf{A}_{jj} + \gamma$.  For all simulations, we set $d=20$.  In the CP0 setting, we select $\gamma \in [0.7, 0.9]$, in CP1 we select $\gamma \in [0.7, 0.9]$ for the first 100 time points, while we select $\gamma \in [0.1, 0.2]$ for the second 100 time points. For CP4 and CP10, the value of $\gamma$ interchanges between $\gamma \in [0.8, 0.9]$ and $\gamma \in [0.05, 0.1]$ in ABABA and ABABABABABA type settings, respectively.



\subsection{Competitor methods} 
To the best of our knowledge, there are no change-point detection methods in the network structure for tensor variate data, hence there is no direct competitor (\citealp{huang2022} and \citealp{qin2023} studied mean changes in tensor-variate data). However, we can compare TenSeg to other methods by performing the following steps.  After applying CP decomposition to the tensor, five competing change-point methods, that detect changes in the covariance structure of multivariate time series data (listed in Table \ref{table:competitors}), are applied to the decomposed data.
\begin{table}
\caption{Competing methods used in the simulation study. \label{table:competitors}}
\begin{center}
\begin{tabular}{@{}lll@{}}
\hline
Method Notation & Reference & {\textsf{R}} package\\
\hline
Sparsified binary segmentation (SBS) & \cite{cho2015} & {\textbf{hdbinseg}}\\
Barnett \& Onnela (BO) & \cite{barnett2016} & -\\
Ratio & \cite{killick2023} & -\\
Wang, Yu, \& Rinaldo (WYR) & \cite{wang2021cpt} & -\\
Galeano \& Pe\~{n}a (GP) & \cite{galeano2007} & -\\
\hline
\end{tabular}
\end{center}
\end{table}
Currently, there is no open software available for implementing the BO, Ratio, WYR, and GP methods, which is also the case for the majority of the methods mentioned in the Introduction.   For BO, \textsf{R} code was provided to us during a previous project.  For the implementation of Ratio, WYR, and GP, relevant code was also used in \cite{killick2023} and the functions can be found in \url{https://github.com/s-ryan1/Covariance_RMT_simulations}.  For SBS, we only present the results for the finer scale of $-1$ used in the wavelet transformation. For TenSeg, after decomposition takes place, we employ CCID with the mean-dominant norm $L_\infty$ combined with the optimization of a model selection criterion, as explained in Section \ref{sec:ccid}.

\subsection{Simulation results}\label{sec:sim_res}
In this section we present and discuss the results obtained from the simulation study carried out for the settings described in Section \ref{subsec:description}. The \textsf{R} code to reproduce the simulation study explained in this section, as well as those of Sections \ref{sec:autocorrelated_data}, \ref{sec:HOSVD_results} and \ref{sec:limitations} can be found at \url{https://github.com/Anastasiou-Andreas/TenSeg}. To study the robustness of TenSeg and competitor methods (and due to the fact that the ALS method in CP is not guaranteed to converge to a global minimum), we varied the number of components, $C_{{\rm CP}}$, in the CP decomposition.  For the simulation study, we ran 500 iterations for each scenario and we present the frequency distribution of $\hat{N} - N$ (comparing the estimated number and the true number of change-points) for each method. In the table results, the methods with the highest empirical frequency of $\hat{N} - N = 0$ and those within $10\%$ of the highest empirical frequency are given in bold. As a measure of the accuracy of the detected locations of the change-points, we provide the scaled Hausdorff distance $d_H = n_s^{-1}\max\left\lbrace \max_j\min_k\left|r_j-\hat{r}_k\right|,\max_k\min_j\left|r_j-\hat{r}_k\right|\right\rbrace,$ where $n_s$ is the length of the largest segment, $r_j$ is the location of the true change-point(s), and $\hat{r}_k$ is the location of the estimated change-point(s).  The objective is to minimize the Hausdorff distance. The average computational time (in seconds) for each method is also provided. 

The results under the AR, ER, and SB precision matrix structures are given in Tables \ref{table:results_AR}, \ref{table:results_ER}, and \ref{table:results_SB}, respectively. Here, we cover the scenarios involving multiple change-points where the detection process is more demanding; results for the at most one change-point case (in any structure) are presented in the Supplementary Materials.

{\setstretch{1}
\begin{table}
\centering
\caption{Simulation results for the AR model with 4 (CP4) and 10 (CP10) true change-points. We used 20, 10, 5 number of components in the CP decomposition. We provide the distribution of $\hat{N} - N$ over 500 simulation iterations, the average Hausdorff distance, $d_H$, and the average computational time (s) for each method.}
\vspace{0.05in}
{\small{
\begin{tabular}{@{}llllllllllll@{}}
\cline{1-12}
 &  &  & \multicolumn{7}{c}{} &  & \\
 &  &  & \multicolumn{7}{c}{$\hat{N} - N$} &  & \\
Method & Model & $C_{{\rm CP}}$ & $\leq -3$ & -2 & -1 &0 & 1 & 2 & $\geq 3$ & $d_H$&  Time (s)\\
\hline
{\bf{TenSeg}} &  &  & 0 & 0 & 0 & {\bf{483}} & 7 & 5 & 5 & 0.006 & 1.633\\
sbs &  &  & 0 & 94 & 0 & 406 & 0 & 0 & 0 & 0.049 & 6.311\\
BO &  &  & 2 & 280 & 6 & 139 & 65 & 7 & 1 & 0.205 & 20.594\\
Ratio & CP4 & 20 & 0 & 3 & 0 & 379 & 105 & 13 & 0 & 0.037 & 0.217\\
WYR &  &  & 0 & 1 & 0 & 332 & 164 & 3 & 0 & 0.037 & 0.217\\
{\bf{GP}} &  &  & 0 & 0 & 0 & {\bf{500}} & 0 & 0 & 0 & 0.001 & 0.029\\
\hline
{\bf{TenSeg}} &  &  & 0 & 0 & 0 & {\bf{495}} & 3 & 2 & 0 & 0.004 & 0.272\\
sbs &  &  & 6 & 257 & 0 & 237 & 0 & 0 & 0 & 0.122 & 2.855\\
BO &  &  & 2 & 81 & 12 & 229 & 145 & 30 & 1 & 0.081 & 18.861\\
Ratio & CP4 & 10 & 0 & 0 & 0 & 180 & 222 & 83 & 15 & 0.054 & 0.254\\
WYR &  &  & 0 & 0 & 0 & 1 & 22 & 343 & 134 & 0.075 & 0.246\\
GP &  &  & 0 & 269 & 0 & 231 & 0 & 0 & 0 & 0.090 & 0.026\\
\hline
{\bf{TenSeg}} &  &  & 0 & 0 & 0 & {\bf{481}} & 17 & 2 & 0 & 0.006 & 0.053\\
sbs &  &  & 76 & 353 & 2 & 69 & 0 & 0 & 0 & 0.291 & 1.985\\
BO &  &  &  5 & 88 & 5 & 210 & 146 & 42 & 3 & 0.084 & 19.652\\
Ratio & CP4 & 5 & 0 & 1 & 0 & 124 & 204 & 123 & 48 & 0.062 & 0.335\\
WYR &  &  & 0 & 0 & 0 & 0 & 0 & 77 & 423 & 0.072 & 0.318\\
GP &  &  &  344 & 0 & 0 & 155 & 1 & 0 & 0 & 0.345 & 0.026\\
\hline
\hline
{\bf{TenSeg}} &  &  & 0 & 0 & 0 & {\bf{489}} & 5 & 1 & 5 & 0.002 & 2.583\\
sbs &  &  & 482 & 18 & 0 & 0 & 0 & 0 & 0 & 0.747 & 8.022\\
BO &  &  & 277 & 14 & 1 & 71 & 83 & 42 & 12 & 0.421 & 82.139\\
Ratio & CP10 & 20 & 0 & 0 & 0 & 74 & 164 & 159 & 103 & 0.037 & 1.057\\
WYR &  &  & 4 & 38 & 15 & 139 & 175 & 98 & 31 & 0.048 & 1.256\\
{\bf{GP}} &  &  & 35 & 0 & 0 & {\bf{465}} & 0 & 0 & 0 & 0.025 & 0.094\\
\hline
{\bf{TenSeg}} &  &  & 0 & 0 & 0 & {\bf{498}} & 2 & 0 & 0 & 0.002 & 0.930\\
sbs &  &  & 500 & 0 & 0 & 0 & 0 & 0 & 0 & 0.909 & 3.918\\
BO &  &  & 152 & 2 & 3 & 120 & 136 & 69 & 18 & 0.138 & 84.425\\
Ratio & CP10 & 10 & 0 & 0 & 0 & 27 & 90 & 133 & 250 & 0.039 & 1.172\\
WYR &  &  & 0 & 0 & 0 & 0 & 0 & 4 & 496 & 0.047 & 1.311\\
GP &  &  & 500 & 0 & 0 & 0 & 0 & 0 & 0 & 0.909 & 0.019\\
\hline
{\bf{TenSeg}} &  &  & 0 & 0 & 0 & {\bf{466}} & 28 & 4 & 2 & 0.004 & 0.172\\
sbs &  &  & 500 & 0 & 0 & 0 & 0 & 0 & 0 & 0.909 & 1.493\\
BO &  &  & 108 & 11 & 4 & 57 & 110 & 122 & 88 & 0.129 & 85.004\\
Ratio & CP10 &  5& 0 & 0 & 0 & 9 & 33 & 89 & 369 & 0.038 & 1.453\\
WYR &  &  & 0 & 0 & 0 & 0 & 0 & 0 & 500 & 0.042 & 1.355\\
GP &  &  & 500 & 0 & 0 & 0 & 0 & 0 & 0 & 0.909 & 0.015\\
\hline
\end{tabular}}}
\label{table:results_AR}
\end{table}
\begin{table}
\centering
\caption{Simulation results for the ER model with 4 (CP4) and 10 (CP10) true change-points.  We used 20, 10, 5 number of components in the CP decomposition. We provide the distribution of $\hat{N} - N$ over 500 simulation iterations, the average Hausdorff distance, $d_H$, and the average computational time (s) for each method.}
\vspace{0.05in}
{\small{
\begin{tabular}{@{}llllllllllll@{}}
\cline{1-12}
 &  &  & \multicolumn{7}{c}{} &  & \\
 &  &  & \multicolumn{7}{c}{$\hat{N} - N$} &  & \\
Method & Model & $C_{{\rm CP}}$ & $\leq -3$ & -2 & -1 &0 & 1 & 2 & $\geq 3$ & $d_H$&  Time (s)\\
\hline
{\bf{TenSeg}} &  &  & 1 & 1 & 7 & {\bf{478}} & 8 & 4 & 1 & 0.019 & 0.641\\
sbs &  &  & 132 & 362 & 6 & 0 & 0 & 0 & 0 & 0.249 & 4.682\\
BO &  &  & 32 & 419 & 46 & 3 & 0 & 0 & 0 & 0.193 & 22.036\\
Ratio & CP4 & 20 & 0 & 2 & 1 & 247 & 219 & 29 & 2 & 0.062 & 0.185\\
WYR &  &  & 1 & 73 & 171 & 171 & 51 & 33 & 0 & 0.305 & 0.174\\
GP &  &  & 223 & 263 & 7 & 7 & 0 & 0 & 0 & 0.338 & 0.021\\
\hline
{\bf{TenSeg}} &  &  & 2 & 9 & 38 & {\bf{439}} & 12 & 0 & 0 & 0.039 & 0.137\\
sbs &  &  & 332 & 161 & 7 & 0 & 0 & 0 & 0 & 0.397 & 1.682\\
BO &  &  & 53 & 331 & 39 & 61 & 13 & 3 & 0 & 0.203 & 15.194\\
Ratio & CP4 & 10 & 0 & 2 & 0 & 284 & 165 & 44 & 5 & 0.045 & 0.232\\
WYR &  &  & 0 & 0 & 43 & 189 & 146 & 58 & 64 & 0.358 & 0.200\\
GP &  &  & 391 & 109 & 0 & 0 & 0 & 0 & 0 & 0.428 & 0.014\\
\hline
{\bf{TenSeg}} &  &  & 0 & 12 & 43 & \textbf{433} & 10 & 2 & 0 & 0.050 & 0.039\\
sbs &  &  & 435 & 65 & 0 & 0 & 0 & 0 & 0 & 0.463 & 1.261\\
BO &  &  & 52 & 271 & 51 & 83 & 26 & 16 & 1 & 0.190 & 16.146\\
Ratio & CP4 & 5 & 0 & 7 & 5 & 243 & 167 & 63 & 15 & 0.053 & 0.253\\
WYR &  &  & 0 & 0 & 0 & 14 & 98 & 134 & 254 & 0.322 & 0.237\\
GP &  &  & 473 & 27 & 0 & 0 & 0 & 0 & 0 & 0.487 & 0.012\\
\hline
\hline
\textbf{TenSeg} &  &  & 0 & 0 & 0 & \textbf{495} & 2 & 2 & 1 & 0.008 & 3.689\\
sbs &  &  & 483 & 16 & 1 & 0 & 0 & 0 & 0 & 0.287 & 7.812\\
BO &  &  & 216 & 162 & 68 & 35 & 18 & 1 & 0 & 0.092 & 80.114\\
Ratio & CP10 & 20 & 0 & 0 & 0 & 24 & 148 & 217 & 111 & 0.036 & 0.746\\
WYR &  &  & 16 & 49 & 60 & 127 & 125 & 81 & 42 & 0.055 & 0.943\\
GP &  &  & 493 & 8 & 0 & 0 & 0 & 0 & 0 & 0.708 & 0.060\\
\hline
\textbf{TenSeg} &  &  & 0 & 0 & 0 & \textbf{492} & 8 & 0 & 0 & 0.009 & 0.661\\
sbs &  &  & 490 & 8 & 2 & 0 & 0 & 0 & 0 & 0.391 & 3.883\\
BO &  &  & 61 & 147 & 53 & 95 & 100 & 39 & 5 & 0.064 & 67.749\\
Ratio & CP10 & 10 & 0 & 0 & 0 & 110 & 196 & 108 & 86 & 0.031 & 0.879\\
WYR &  &  & 0 & 0 & 0 & 1 & 0 & 6 & 493 & 0.056 & 1.014\\
GP &  &  & 500 & 0 & 0 & 0 & 0 & 0 & 0 & 0.890 & 0.022\\
\hline
\textbf{TenSeg} &  &  & 0 & 1 & 6 & \textbf{441} & 43 & 8 & 1 & 0.018 & 0.128\\
sbs &  &  & 499 & 1 & 0 & 0 & 0 & 0 & 0 & 0.612 & 1.342\\
BO &  &  & 93 & 64 & 41 & 40 & 148 & 73 & 41 & 0.093 & 63.978\\
Ratio & CP10 & 5 & 0 & 0 & 0 & 54 & 129 & 134 & 183 & 0.033 & 1.029\\
WYR &  &  & 0 & 0 & 0 & 0 & 0 & 0 & 500 & 0.064 & 1.059\\
GP &  &  & 500 & 0 & 0 & 0 & 0 & 0 & 0 & 0.907 & 0.015\\
\hline
\end{tabular}}}
\label{table:results_ER}
\end{table}
\begin{table}
\centering
\caption{Simulation results for the SB model with 4 (CP4) and 10 (CP10) true change-points.  We used 20, 10, 5 number of components in the CP decomposition. We provide the distribution of $\hat{N} - N$ over 500 simulation iterations, the average Hausdorff distance, $d_H$, and the average computational time (s) for each method.}
\vspace{0.05in}
{\small{
\begin{tabular}{@{}llllllllllll@{}}
\cline{1-12}
 &  &  & \multicolumn{7}{c}{} &  & \\
 &  &  & \multicolumn{7}{c}{$\hat{N} - N$} &  & \\
Method & Model & $C_{{\rm CP}}$ & $\leq -3$ & -2 & -1 &0 & 1 & 2 & $\geq 3$ & $d_H$&  Time (s)\\
\hline
\textbf{TenSeg} &  &  & 0 & 0 & 0 & \textbf{484} & 4 & 9 & 3 & 0.006 & 1.842\\
sbs &  &  & 0 & 443 & 0 & 57 & 0 & 0 & 0 & 0.171 & 4.216\\
BO &  &  & 6 & 414 & 25 & 19 & 29 & 7 & 0 & 0.179 & 19.879\\
Ratio & CP4 & 20 & 0 & 0 & 1 & 400 & 95 & 4 & 0 & 0.030 & 0.193\\
WYR &  &  & 2 & 71 & 206 & 96 & 81 & 39 & 5 & 0.198 & 0.225\\
{\textbf{GP}} &  &  & 0 & 0 & 0 & {\textbf{500}} & 0 & 0 & 0 & 0.001 & 0.029\\
\hline
\textbf{TenSeg} &  &  & 0 & 0 & 0 & \textbf{493} & 4 & 2 & 1 & 0.004 & 0.247\\
sbs &  &  & 24 & 461 & 0 & 15 & 0 & 0 & 0 & 0.215 & 2.001\\
BO &  &  & 0 & 220 & 8 & 154 & 93 & 18 & 7 & 0.106 & 18.892\\
Ratio & CP4 & 10 & 0 & 0 & 0 & 147 & 217 & 108 & 28 & 0.061 & 0.238\\
WYR &  &  & 0 & 0 & 0 & 5 & 12 & 17 & 466 & 0.237 & 0.297\\
GP &  &  & 37 & 195 & 0 & 268 & 0 & 0 & 0 & 0.194 & 0.018\\
\hline
\textbf{TenSeg} &  &  & 0 & 0 & 0 & \textbf{471} & 28 & 1 & 0 & 0.007 & 0.055\\
sbs &  &  & 271 & 223 & 0 & 6 & 0 & 0 & 0 & 0.432 & 1.345\\
BO &  &  & 14  & 62 & 5 & 237 & 133 & 44 & 5 & 0.077 & 18.825\\
Ratio & CP4 & 5 & 0 & 1 & 0 & 83 & 176 & 142 & 98 & 0.075 & 0.264\\
WYR &  &  & 0 & 0 & 0 & 0 & 0 & 0 & 500 & 0.263 & 0.328\\
GP &  &  &  495 & 0 & 0 & 5 & 0 & 0 & 0 & 0.815 & 0.009\\
\hline
\hline
\textbf{TenSeg} &  &  & 0 & 0 & 0 & \textbf{498} & 0 & 0 & 2 & 0.002 & 8.065\\
sbs &  &  & 500 & 0 & 0 & 0 & 0 & 0 & 0 & 0.362 & 8.212\\
BO &  &  & 463 & 4 & 0 & 11 & 14 & 8 & 0 & 0.332 & 52.993\\
Ratio & CP10 & 20 & 0 & 0 & 0 & 106 & 159 & 156 & 79 & 0.034 & 0.782\\
WYR &  &  & 40 & 31 & 23 & 78 & 119 & 107 & 102 & 0.066 & 1.221\\
GP &  &  & 445 & 0 & 0 & 55 & 0 & 0 & 0 & 0.808 & 0.086\\
\hline
\textbf{TenSeg} &  &  & 0 & 0 & 0 & \textbf{499} & 1 & 0 & 0 & 0.002 & 1.043\\
sbs &  &  & 500 & 0 & 0 & 0 & 0 & 0 & 0 & 0.472 & 3.432\\
BO &  &  & 82 & 3 & 0 & 231 & 98 & 63 & 23 & 0.061 & 90.179\\
Ratio & CP10 & 10 & 0 & 0 & 0 & 19 & 89 & 135 & 257 & 0.040 & 1.005\\
WYR &  &  & 0 & 0 & 0 & 0 & 0 & 0 & 500 & 0.054 & 1.421\\
GP &  &  & 500 & 0 & 0 & 0 & 0 & 0 & 0 & 0.909 & 0.014\\
\hline
\textbf{TenSeg} &  &  & 0 & 0 & 0 & \textbf{457} & 36 & 7 & 0 & 0.006 & 0.174\\
sbs &  &  & 500 & 0 & 0 & 0 & 0 & 0 & 0 & 0.643 & 1.492\\
BO &  &  & 74 & 0 & 4 & 108 & 157 & 89 & 68 & 0.071 & 84.023\\
Ratio & CP10 & 5 & 0 & 0 & 0 & 8 & 43 & 72 & 377 & 0.039 & 1.473\\
WYR &  &  & 0 & 0 & 0 & 0 & 0 & 0 & 500 & 0.064 & 1.439\\
GP &  &  & 500 & 0 & 0 & 0 & 0 & 0 & 0 & 0.909 & 0.013\\
\hline
\end{tabular}}}
\label{table:results_SB}
\end{table}}
Across the 18 combinations of multiple change-points (CP4 and CP10) and tensor decomposition components ($C_{{\rm CP}} = 20, 10, 5$), TenSeg performs best in 16 in terms of the frequency distribution of $\hat{N} - N$ (comparing the estimated number and the true number of change-points) across the 500 iterations.  In the two combinations that the GP algorithm performs best, TenSeg is second best and is within at most 17 out of the 500 iterations in terms of the frequency of $\hat{N} = N$.  However, GP performs poorly in the other settings.  In the other 16 combinations where TenSeg has the best performance, the performance of competitors is significantly inferior to that of TenSeg. The good performance of TenSeg in terms of $\hat{N} - N$ is uniform across simulations with a contrasting number of change-points, different structures (AR, ER, and SB) for the precision matrix, and a different number of components employed for the CP decomposition. TenSeg completely dominates all other competing state-of-the-art methods in terms of the number of iterations that $\hat{N} = N$; in most cases, this number is greater than 450 (out of 500).  None of the other competing methods has as stable results compared to TenSeg.  For example, GP performs best for the CP4 (4 change-points) with $C_{{\rm CP}} = 20$ (20 components in the CP decomposition) for the AR and SB models, but its performance deteriorates considerably as the number of true change-points increases and especially when the number of components in the CP decomposition decreases. This pattern is also evident in all competitors except for BO, which obtains a stable performance as the number of components varies, but overall produces inferior results across the simulations. Furthermore, the competing methods do not exhibit a uniform performance under the different precision matrices (see, for example, GP's performance for the ER model). However, TenSeg's performance remains stable across all combinations. For example, for the difficult case of CP10 (10 change-points) in the AR case with $C_{{\rm CP}} = 5$ (5 components in the CP decomposition), TenSeg has $\hat{N} = N$ in 466/500 iterations, while the next closest competitor achieves $\hat{N} = N$ in only 57 iterations (Table \ref{table:results_AR}).  TenSeg performs particularly well in the 10 true change-point scenario for all $C_{{\rm CP}}$ ($= 20, 10, 5$) combinations.

In terms of the accuracy of the estimated change-point locations, as measured by the Hausdorff distance, TenSeg has the minimum Hausdorff distance in 16 of the 18 combinations of multiple change-points and tensor decomposition components. Again, similar to the frequency distribution of $\hat{N} - N$ above, in the two combinations that GP performs best, TenSeg has a Hausdorff distance of $d_{H} = 0.006$, while GP has $d_{H} = 0.001$.  However, GP's performance across all the other simulations is not consistent. In most cases, TenSeg's $d_{H}$ is at least one order of magnitude less than that of the competing methods.  

While TenSeg has the best performance in terms of estimating the true number of multiple change points and in terms of the accuracy of the location of the estimated change-points, it also performs well in terms of computational speed.  TenSeg's low average computational time in the simulations provides evidence that the algorithm can efficiently detect the change-points in the structure of tensor objects with at least 4 modes of extent in the range of tens or even hundreds; this is completed in the order of seconds using a conventional processing machine. In addition to accuracy and computational efficiency, TenSeg has impressive robust behavior with respect to the number of components used in the CP decomposition method; more precisely, the method shows no signs of performance degradation with a reduction in the number of decomposition components. As shown in Section \ref{sec:HOSVD_results}, TenSeg is also robust in the choice of the decomposition method utilized. Combining the aforementioned conclusions with the results presented in the Supplementary Materials covering cases of at most one change-point, we conclude that TenSeg is an accurate, robust, reliable, and computationally efficient method for the detection of changes in the network structure of tensors. 

\section{Real data} \label{sec:real_data_results}
\subsection{Ethereum blockchain results}
We used the NORMO estimator (Section \ref{sec:normo}) to find the optimal rank in the CP decomposition of the data set, which selected a rank of 4 components.  We then applied our TenSeg method: we chose the number of change-points using the model selection approach (see Section \ref{sec:ccid}; for all the particular details on the information criterion using a penalty function and the pseudo-likelihood of the given data, please refer to Section 2.7 of \citealp{anastasiou2022}). The detected change-points were at time points $t = $ 31, 66, 98, 115, 160, 209, 258.  As the true change-points are unknown for this data set, we attempt to coincide the change-points detected by TenSeg with macro Ethereum blockchain events from Wikipedia (\url{https://timelines.issarice.com/wiki/Timeline_of_Ethereum}), which lists and explains the major Ethereum blockchain events since 2015.
The 7 aforementioned change-points detected by TenSeg correspond to the following dates:
\begin{itemize}
\item August 11, 2017 (change-point at $t = 31$): On August 12, 2017, Microsoft released the Confidential Consortium Framework, an Ethereum-based protocol that allowed commercial companies and other large-scale organizations to process information on the Ethereum blockchain with increased privacy.
\item  September 15, 2017 (change-point at $t=66$): This change-point is possibly related to two events that occurred close to the detection time. First, on September 4, 2017, the Chinese regulator ordered all initial coin offerings to be halted in the country, for exchanges to stop trading in tokens, and for funds already raised to be returned to investors. This decision prompted the Ethereum price to plunge further. Second, on September 8, 2017, the Blackmoon Financial Group announced the launch of the Blackmoon Crypto Platform, an Ethereum blockchain-based platform for tokenized vehicles.
\item October 17, 2017 (change-point at $t = 98$): On October 16, 2017, the Byzantium hard fork (a protocol upgrade) occurred at block number 4,370,000. Planned as the first phase of the next stage of Ethereum's roadmap known as Metropolis, this fork aimed to reduce the block mining rewards from 5 to 3 ETH, to delay the difficulty bomb by a year, to add the ability to make non-state-changing calls to other contracts, and to provide certain cryptography methods that allow for layer 2 scaling.
\item November 3, 2017 (change-point at $t = 115$): Two important events occurred near this estimated change-point location. First, on October 31, Amazon Technologies Inc., which is a subsidiary of Amazon, registered another website related to Ethereum  (\url{amazonethereum.com}), which was added to \url{amazoncryptocurrency.com} and \url{amazoncryptocurrencies.com}. Second, on November 6, a major security breach occurred when a GitHub user managed to exploit a vulnerability within the smart-contract library code of the Parity wallet service, blocking funds in 587 wallets that contained 513,774.16 Ether as well as various other tokens. This led to the freezing of hundreds of millions of dollars in Ethereum.
\item  December 18, 2017 (change-point at $t = 160$): On  January 4, 2018, the Ethereum valuation exceeded US\$1,000 for the first time.
\item February 5, 2018 (change-point at $t = 209$): This estimated change-point is associated with two events. First, on February 2, 2018, Vitalik Buterin, co-founder of Ethereum, donated US\$2.4 million of Ethereum to the SENS Research Foundation, which is a non-profit organization that conducts research programs and public relations work for the application of regenerative medicine to aging. Second, on February 5, 2018, the same date as the estimated location, UNICEF launched the Game Chaingers scheme, a project to recruit gamers to use the processing power of their computers to mine Ethereum to raise money for Syrian children.
\item March 26, 2018 (change-point at $t = 258$): This change-point is possibly related to the event in March 2018, when the Ethereum Foundation announced the first wave of grants, with 13 projects receiving a combined US\$2.5 million.
\end{itemize} 

\begin{figure}
\begin{center}
\includegraphics[height=3.5cm, width=13cm]{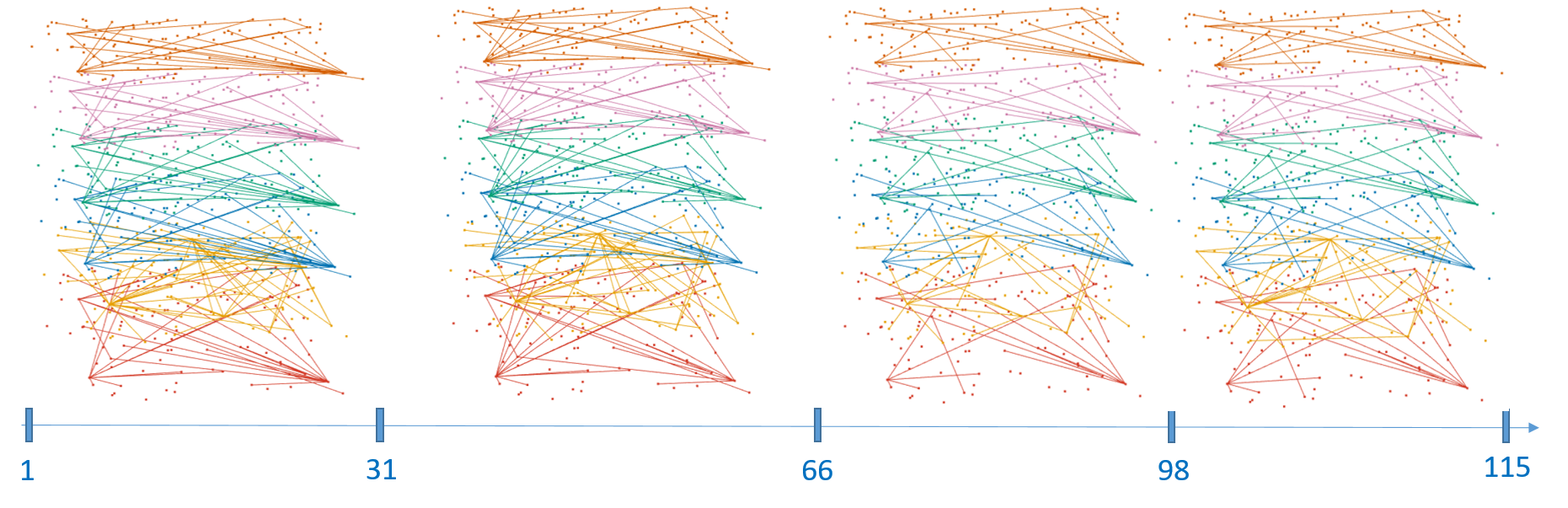}
\includegraphics[height=3.5cm, width=13cm]{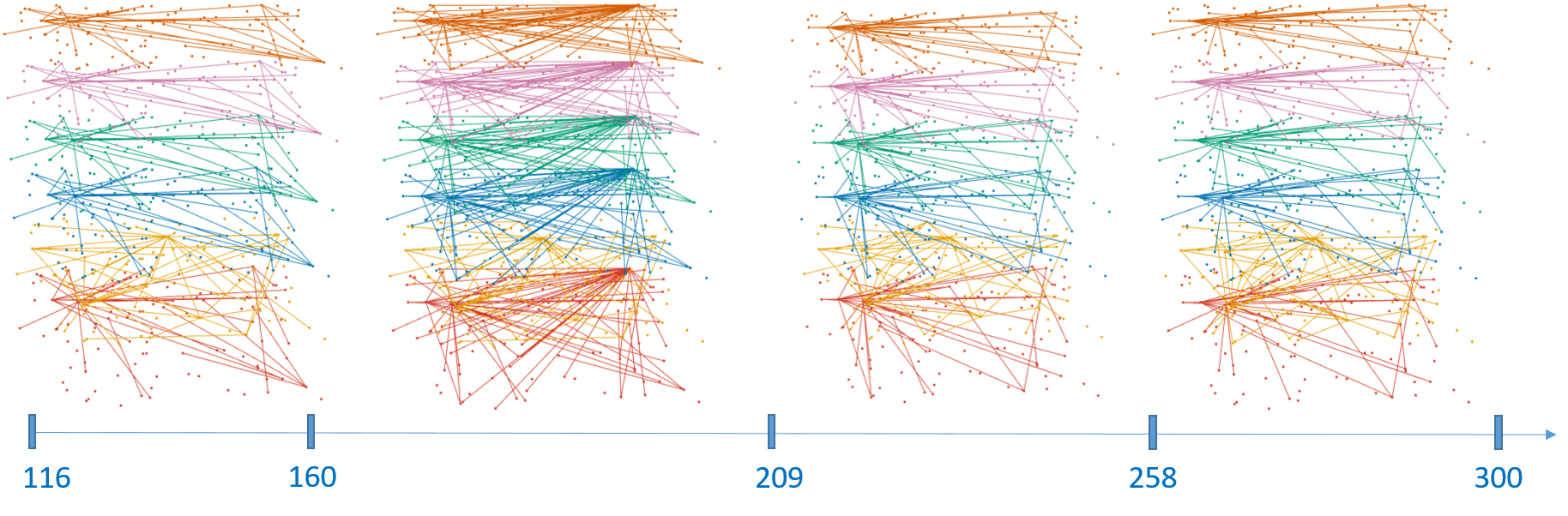}
\caption{The activity networks between the 100 investors for the 6 tokens namely “bat”, “golem”, “mcap”, “statusnetwork”, “storj”, and “tenxpay” between the detected change-points at $t = 31, 66, 98, 115, 160, 209, 258$.}
\label{fig:eth_network}
\end{center}
\end{figure}

Figure \ref{fig:eth_network} shows the activity networks between the 100 investors for the 6 tokens, namely ``bat'', ``golem'', ``mcap'', ``statusnetwork'', ``storj'', and ``tenxpay'', between the detected change-points.  Each token is represented by a different network color, and each node in each token layer network represents an investor (the location of each investor remains constant throughout the time course).  It is evident that the networks are changing over time and that the level of activity increases and decreases between stationary networks.  Some investors are very active during some time periods but become inactive during others.  The third stationary network (between time points $t=66$ and $t=98$ in Figure \ref{fig:eth_network}) is quite sparse compared to the other stationary networks, which is consistent with the event on September 4, 2017, when the Chinese regulator ordered all initial coin offerings to be stopped in the country, for exchanges to stop trading in existing tokens, and for funds already raised to be returned to investors. This decision prompted the Ethereum price to plunge subsequently, and in our network, the event had the effect of decreasing the activity between the investors.  The sixth stationary network (between time points $t=160$ and $t=209$ in Figure \ref{fig:eth_network}) is very dense compared to the other stationary networks, which coincides with the event on January 4, 2018, when the Ethereum valuation surpassed US\$1,000 for the first time, creating a buzz and thereby increasing the activity between investors.  There is a particular investor node that has a great deal of activity with the other investors across all tokens during this stationary network.  This surge in activity is notable and may indicate potential market manipulation, which should have been further investigated.  It is such events that show the capability of TenSeg as we can observe changes across several trading networks or a ``network of networks'' over time, instead of only considering changes in one network.  Market manipulators are now trading on multiple cryptocurrency platforms, making their detection more difficult. TenSeg is an important contribution to this space.

To explore the robustness of the change-points detected for the Ethereum data set, we also applied TenSeg with the thresholding stopping criterion.  In this case, we obtain the change-points at time points $t=33, 51, 98, 115, 160, 209, 258$, which has a significant overlap with the change-points detected when TenSeg was combined with the information criterion rule.  Without TenSeg, the analysis of this data set would not be possible in its current form.  Currently, the only course of action would be to analyze each token network separately and then combine the results across networks either quantitatively or qualitatively. 

\subsection{Enron data} \label{subsec:enron}
To show the generalizability of TenSeg, we also applied it to the Carnegie Mellon University CALO Project Enron email dataset. The CALO project contains data from about 150 users and their email exchanges. For more information, see \url{https://enrondata.readthedocs.io/en/latest/data/calo-enron-email-dataset/}.  In this data set, there are 184 unique email addresses followed for over 189 weeks from November 1998 to June 2002 and the content of the emails is categorized based on 33 different topics. The data are available in \url{https://www.cis.jhu.edu/~parky/Enron/}.  For this data set, we consider the sender and receiver of emails over 189 weeks on the 33 topics.  Therefore, the tensor object that we created for this data set is of dimension $33 \times 184 \times 184 \times 189$ (topics $\times$ network $\times$ time).  Our objective is to study whether the patterns of email exchange change across the topics jointly. Each email address is regarded as one subject or a node in the networks. The adjacency matrices of the dynamic networks,  $A_t, t = 1, 2, \ldots, 189$, are of dimension $184 \times 184$ and reflect the links between any two subjects. If there is an email from subject $i$ to subject $j$ at time point $t$, then the ($i, j$) element of $A_t$
is $1$; otherwise, it is zero.  This data set has been analyzed in several academic articles \citep{Priebe2006,DeRidder2016,Luo2023} from a change-point perspective but only on a single topic.  It has never been analyzed from the tensor network change-point perspective, where all 33 topics can be considered jointly.  The results from applying TenSeg can be found in the Supplementary Materials and contain new perspectives on the data set. 

%
%
%
%
%
%


\section{Discussion} \label{sec:discussion}

\subsection{Impact of autocorrelation} \label{sec:autocorrelated_data}
All the simulations carried out in Section \ref{sec:sims} assume that the data are independent and identically distributed across time.  As many other articles have discussed, the addition of temporal dependence makes the identification of change-points far more challenging. In this section, the simulations in Section \ref{subsec:description} are extended to serially correlated data. More specifically, we work in the scenario in which the noise in the simulated data is generated from an AR(1) process with a correlation coefficient set to $\alpha = 0.7$. Table 5 in the Supplementary Materials presents the results for three different precision matrices, AR, SB, and ER (see Section \ref{sec:sims} for an explanation).  For the AR and SB random graph models combined with autocorrelation, we deduce that TenSeg preserves its behavior in terms of accuracy for both the estimated number and the locations of the change-points.  For the simulated ER random graph models combined with autocorrelation, TenSeg performs well in the at most one change-point scenario; however, its performance begins to deteriorate in the multiple change-point cases.  This behavior is expected given the difficulty of detecting change-points in networks created by ER type random graph models compared to random graph models with a particular structure. In ER graph models, edges follow a general structure with edges often isolated, thereby making change-points difficult to detect (especially in the presence of autocorrelation); see \cite{zhu2018} for more details.

Heuristically, the stable behaviour of TenSeg, even in the presence of serial correlation, is expected, as TenSeg relies heavily on the isolation of each one of the change-points before their detection, that is, the change-points are detected in intervals that contain no other change-points; such scenarios enhance the detection power in general. Practical ways that can be used to further improve TenSeg’s performance in the presence of high serial correlation in data are subsampling and pre-averaging techniques, which have already been described in Section 5.1 of \cite{anastasiou2022}. Briefly, subsampling first requires choosing a positive integer $s$ and then subsamples every $s$ data points from the original data; this creates $s$ mutually exclusive (sub-sampled) data sequences, in which the autocorrelation is expected to be reduced compared to the original data. After these subsampled data sequences have been created, TenSeg can be applied to each one of them and return $s$ different sets of estimated change-points. A majority voting rule can then be applied to these obtained sets and only those values that appear at least a pre-decided number of times are retained.  Regarding the pre-averaging technique, the data are uniformly averaged over prespecified short time periods, which can also significantly reduce
autocorrelation.

\subsection{Robustness to choice of decomposition}
\label{sec:HOSVD_results}
In this section, the robustness of TenSeg is investigated by considering a different decomposition method. To be more precise, we employ the Higher-Order Singular Value Decomposition (HOSVD) method for decomposing the initial tensor, instead of the CP decomposition method used in Section \ref{sec:sim_res}. HOSVD is a tensor factorization technique that is used to decompose a high-dimensional tensor into a set of core tensors and mode matrices. It is a generalization of the Singular Value Decomposition (SVD) for matrices to higher-dimensional tensors.  Given a tensor $\mathcal{X} \in \mathbb{R}^{n_1 \times n_2 \times \ldots \times n_N}$ of order $N$, the HOSVD decomposes it into the following components:
a core tensor $\mathcal{G} \in \mathbb{R}^{R_1 \times R_2 \times \ldots \times R_N}$ and mode matrices $\mathbf{U}_1, \mathbf{U}_2, \ldots, \mathbf{U}_N$.  The core tensor captures the most significant information in the original tensor. The dimensions $R_i$ determine the rank of the decomposition and represent the reduced dimensions along each mode; therefore, it holds that $R_i \leq n_i, \forall i = 1,2,\ldots, N$. Mode matrices are orthogonal matrices for each mode of the original tensor. They contain the basis vectors that span the subspace along each mode.  HOSVD is obtained by computing the mode-wise matrix unfoldings of the original tensor and applying SVD separately to each mode. Mathematically, the decomposition can be expressed as
$\mathcal{X} = \mathcal{G} \times_1 \mathbf{U}_1 \times_2 \mathbf{U}_2 \times_3 \ldots \times_N \mathbf{U}_N,$ where $\times_i$ represents the mode-$i$ tensor multiplication.

For uniformity with the results in Section \ref{sec:sim_res}, we use the same number of components, that is, 5, 10, and 20. The performance of TenSeg using the HOSVD decomposition in simulations in which the precision matrices undergo changes in an AR, SB, or ER structure is shown in Table 6 of the Supplementary Materials.  Similar to the presence of autocorrelation in the data, apart from a slight underestimation of the number of change-points in the case of ER changes with multiple change-points and for the number of decomposition components being relatively low (equal to 5), we find that TenSeg preserves its accurate behavior with HOSVD. Therefore, we conclude that TenSeg's performance appears to remain unaffected under different choices of decomposition methods.

\subsection{Limitations}
\label{sec:limitations}
In this section, our objective is to test the limits of TenSeg, hence we apply it to simulated data where the magnitudes of the changes are small, close to negligible.  More specifically, we recreate the AR1($\rho$) random graph model in Section \ref{subsec:description}. Again, the covariance matrix is of the form $\mathbf{A} = (\rho^{|i-j|})_{ij}$.  However, we make changes of smaller magnitude.  In the CP1 setting, we take $\rho = 0.4$ (instead of $0.2$ used in Section \ref{sec:sim_res}) for the first 100 time points, while $\rho = 0.6$ (instead of $0.8$ used in Section \ref{sec:sim_res}) for the second 100 time points. For CP4 and CP10, the value of $\rho$ alternates between 0.4 and 0.6 in ABABA and ABABABABABA type settings, respectively. Therefore, in this section, the magnitude of the changes in the parameter $\rho$ is equal to 0.2 (which is quite low), while in Section \ref{sec:sim_res}, Table \ref{table:results_AR}, the results shown are when the magnitude of the changes is equal to 0.6. We highlight that results are presented for different values of the number of components, $C_{{\rm CP}}$, used in the CP decomposition. As in previous sections, for the simulation study, we ran 500 iterations for each setting and, as a measure of the accuracy of the estimated change-point number, we present the frequency distribution of $\hat{N} - N$ for each method. The methods with the highest empirical frequency of $\hat{N} - N = 0$ and those within $10\%$ off the highest empirical frequency are presented in bold. Furthermore, the scaled Hausdorff distance is provided in the tables as a measure of the accuracy of the detected locations of the change-points. The results for CP4 and CP10 are given in Table \ref{table:results_AR_diff}, while for CP1, the results can be found in the Supplementary Materials.

In terms of accuracy in both the estimated number and the estimated locations of change-points, TenSeg performs, although not as well as in the results of Section \ref{sec:sim_res}, best in all combinations presented in Table \ref{table:results_AR_diff}. The other competitor methods perform poorly.  It is surprising that even in such a difficult structure TenSeg manages to detect the correct number of change-points (in both CP4 and CP10) in at least 75\% of the cases as long as the number of components used in the CP decomposition is sufficiently large. However, in contrast to the large deviation scenarios covered in Section \ref{sec:sim_res}, there are apparent signs of performance degradation as we decrease the number of components used in CP. This leads to insufficient performance for TenSeg, especially in the CP10 setting, when the number of components is equal to 5. Finally, we highlight that TenSeg retains its low average computational time in these difficult structures with small changes in magnitudes.  As in Section \ref{sec:sim_res}, the \textsf{R} code to replicate the simulation study explained in this section can be found in \url{https://github.com/Anastasiou-Andreas/TenSeg}. 
\begin{table}
\centering
\caption{Simulation results for the AR model with smaller change magnitudes (equal to 0.2, instead of 0.6 used in Table \ref{table:results_AR}) when we have 4 (CP4) and 10 (CP10) true change-points. We used 20, 10, 5 number of components in the CP decomposition. We provide the distribution of $\hat{N} - N$ over 500 simulation iterations, the average Hausdorff distance, $d_H$, and the average computational time (s) for each method.}
\vspace{0.05in}
{\small{
\begin{tabular}{@{}llllllllllll@{}}
\cline{1-12}
 &  &  & \multicolumn{7}{c}{} &  & \\
 &  &  & \multicolumn{7}{c}{$\hat{N} - N$} &  & \\
Method & Model & $C_{{\rm CP}}$ & $\leq -3$ & -2 & -1 &0 & 1 & 2 & $\geq 3$ & $d_H$&  Time (s)\\
\hline
\textbf{TenSeg} &  &  & 0 & 0 & 36 & \textbf{446} & 12 & 6 & 0 & 0.032 & 0.478\\
sbs &  &  & 310 & 190 & 0 & 0 & 0 & 0 & 0 & 0.630 & 6.873\\
BO &  &  & 141 & 354 & 0 & 0 & 5 & 0 & 0 & 0.464 & 10.988\\
Ratio & CP4 & 20 & 0 & 352 & 55 & 29 & 54 & 10 & 0 & 0.158 & 0.202\\
WYR &  &  & 6 & 43 & 7 & 300 & 139 & 5 & 0 & 0.084 & 0.212\\
GP &  &  & 2 & 435 & 0 & 63 & 0 & 0 & 0 & 0.207 & 0.024\\
\hline
\textbf{TenSeg} &  &  & 0 & 29 & 24 & \textbf{442} & 4 & 1 & 0 & 0.043 & 0.095\\
sbs &  &  & 460 & 40 & 0 & 0 & 0 & 0 & 0 & 0.786 & 3.052\\
BO &  &  & 75 & 386 & 14 & 10 & 15 & 0 & 0 & 0.378 & 8.499\\
Ratio & CP4 & 10 & 0 & 221 & 69 & 98 & 88 & 23 & 1 & 0.132 & 0.202\\
WYR &  &  & 0 & 0 & 0 & 10 & 56 & 328 & 106 & 0.072 & 0.219\\
GP &  &  & 488 & 12 & 0 & 0 & 0 & 0 & 0 & 0.497 & 0.015\\
\hline
\textbf{TenSeg} &  &  & 7 & 81 & 91 & \textbf{307} & 14 & 0 & 0 & 0.107 & 0.022\\
sbs &  &  & 487 & 13 & 0 & 0 & 0 & 0 & 0 & 0.806 & 1.887\\
BO &  &  & 51 & 289 & 45 & 65 & 37 & 13 & 0 & 0.278 & 8.731\\
Ratio & CP4 & 5 & 0 & 102 & 43 & 147 & 146 & 51 & 11 & 0.097 & 0.252\\
WYR &  &  & 0 & 0 & 0 & 0 & 2 & 61 & 437 & 0.072 & 0.267\\
GP &  &  &  500 & 0 & 0 & 0 & 0 & 0 & 0 & 0.834 & 0.010\\
\hline
\hline
\textbf{TenSeg} &  &  & 2 & 36 & 34 & \textbf{391} & 30 & 5 & 2 & 0.031 & 1.486\\
sbs &  &  & 500 & 0 & 0 & 0 & 0 & 0 & 0 & 0.909 & 14.246\\
BO &  &  & 500 & 0 & 0 & 0 & 0 & 0 & 0 & 0.909 & 11.650\\
Ratio & CP10 & 20 & 404 & 38 & 26 & 17 & 9 & 2 & 4 & 0.220 & 0.798\\
WYR &  &  & 474 & 6 & 2 & 13 & 3 & 2 & 0 & 0.680 & 0.426\\
GP &  &  & 500 & 0 & 0 & 0 & 0 & 0 & 0 & 0.909 & 0.018\\
\hline
\textbf{TenSeg} &  &  & 47 & 73 & 76 & \textbf{273} & 27 & 4 & 0 & 0.075 & 0.269\\
sbs &  &  & 500 & 0 & 0 & 0 & 0 & 0 & 0 & 0.909 & 5.062\\
BO &  &  & 500 & 0 & 0 & 0 & 0 & 0 & 0 & 0.904 & 8.334\\
Ratio & CP10 & 10 & 450 & 12 & 11 & 9 & 10 & 6 & 2 & 0.062 & 1.186\\
WYR &  &  & 4 & 2 & 5 & 9 & 11 & 47 & 422 & 0.062 & 1.186\\
GP &  &  & 500 & 0 & 0 & 0 & 0 & 0 & 0 & 0.909 & 0.015\\
\hline
\textbf{TenSeg} &  &  & 326 & 84 & 41 & \textbf{49} & 0 & 0 & 0 & 0.185 & 0.058\\
sbs &  &  & 500 & 0 & 0 & 0 & 0 & 0 & 0 & 0.909 & 2.403\\
BO &  &  & 500 & 0 & 0 & 0 & 0 & 0 & 0 & 0.853 & 10.415\\
Ratio & CP10 & 5 & 322 & 40 & 25 & 29 & 27 & 28 & 29 & 0.218 & 1.021\\
WYR &  &  & 0 & 0 & 0 & 0 & 0 & 0 & 500 & 0.042 & 1.324\\
GP &  &  & 500 & 0 & 0 & 0 & 0 & 0 & 0 & 0.909 & 0.016\\
\hline
\end{tabular}}}
\label{table:results_AR_diff}
\end{table}
\section{Conclusion} \label{sec:conc}
In this study, we developed a new method, called Tensor Segmentation (TenSeg), to detect multiple change-points in tensors.  We assumed that the true number and the locations of the change-points are unknown a priori. TenSeg first employs a tensor decomposition and then detects multiple change-points in the second-order (cross-covariance or network) structure of the decomposed data. To detect the change-points in the decomposed data, TenSeg then uses a scaled CUSUM statistic and aggregates across the multivariate time series using either the $\ell_\infty$ or the $\ell_2$ norm as explained in Section \ref{sec:ccid}. TenSeg employed the Isolate-Detect (ID) principle to find the multiple change-points. ID works by isolating each of the true change-points within subintervals prior to their detection.  We apply TenSeg to a set of networks from the Ethereum blockchain.  TenSeg assumed a tensor-variate structure of the data, since crime on blockchain transaction networks, such as money laundering or marlet manipulation, usually involves multiple parties who have the opportunity to move funds across multiple cryptocurrency ledgers.  TenSeg detected 7 change-points that coincided with macro blockchain events, but also showed unusual activity at the micro investor level.

We showed, using an extensive simulation study, that TenSeg generally had significantly higher statistical power in various challenging scenarios compared to state-of-the-art change-point methodologies that had to be adapted to a tensor setting.  We showed that TenSeg is also computationally efficient and has reasonable speed compared to competing methods. Computational costs may be challenging at higher dimensions, but as the first step of our method requires a decomposition, higher dimensions are still possible.

TenSeg is novel as it allows for accurate detection in the presence of frequent changes of possibly small magnitudes, and is computationally fast. While TenSeg was applied to Ethereum cryptocurrency data,  we also applied it to the Carnegie Mellon University CALO Project Enron email dataset, showing its generalizability. Furthermore, TenSeg has been shown to be robust in the employment of another decomposition method (HOSVD) as well as in the presence of autocorrelation in the data. To conclude, TenSeg pertains to a general setting and can also be used in a variety of situations where one wishes to study the evolution of a high-dimensional tensor network over time.

\newpage
\bibliographystyle{agsm}
{\setstretch{1}
\bibliography{references}}

\newpage

  \bigskip
  \bigskip
  \bigskip
\begin{center}
    {\LARGE\bf Supplementary Materials for ``Tensor time series change-point detection in cryptocurrency network data''}
\end{center}
  \medskip

\bigskip

We provide supplementary material for the article {\em Tensor time series change-point detection in cryptocurrency network data}. Specifically, Section \ref{sec:further_sim} provides an investigation of the performance of our proposed method, TenSeg, and the competing methods under the no change-point (CP0) and one change-point (CP1) scenarios.  In Section \ref{sec:res_enron}, the results of the application of TenSeg to the Carnegie Mellon University CALO Project Enron email dataset as discussed in Section 5.2 of the main article are presented. In Section \ref{sec:res_auto_HOSVD}, we provide the tables of the results discussed in Sections 6.1, 6.2, 6.3 of the main article, which focus on the performance of our TenSeg method on autocorrelated data, when the HOSVD (instead of the CP) decomposition method was employed, and the limitations of the TenSeg method, respectively.



\setcounter{section}{0}

\section{Further simulation results}
\label{sec:further_sim}
In this section, we investigate the performance of TenSeg and the competing state-of-the-art methods, namely sbs, BO, Ratio, WYR, and GP (see Table 1 of the main article for the relevant references regarding the competitors) under the no change-point (CP0) and one change-point (CP1) scenarios for the AR, ER, and SB structures as described in detail in Section 4.1 of the main article.  Tables \ref{table:results_AR_app}, \ref{table:results_ER_app}, and \ref{table:results_SB_app} show the results for the AR, ER, and SB structures, respectively. In addition, Table \ref{table:results_AR_app_diff} presents the results of small change magnitudes for the CP1 setting as explained in Section 6.3 of the main article. 

The results indicate that in the relevantly simple aforementioned settings, almost all methods exhibit a very good behavior in accurately estimating no change-points (CP0 setting) and the true number and the location of the change-point (CP1 setting). The only method that appears to struggle is WYR, which tends to mostly overestimate the true number of change-points.  As mentioned in the main article, the \textsf{R} code to reproduce the results presented in this section can be found at \url{https://github.com/Anastasiou-Andreas/TenSeg}.
{\setstretch{1}
\begin{table}
\centering
\caption{Simulation results for the AR model with 0 (CP0) and 1 (CP1) true change-points.  We used 20, 10, 5 number of components in the CP decomposition. We provide the distribution of $\hat{N} - N$ over 500 simulation iterations, the average Hausdorff distance, $d_H$, and the average computational time (s) for each method.}
\vspace{0.05in}
{\small{
\begin{tabular}{|l|l|l|l|l|l|l|l|l|l|l|l|}
\cline{1-12}
 &  &  & \multicolumn{7}{|c|}{} &  & \\
 &  &  & \multicolumn{7}{|c|}{$\hat{N} - N$} &  & \\
Method & Model & $C_{{\rm CP}}$ & $\leq -3$ & -2 & -1 &0 & 1 & 2 & $\geq 3$ & $d_H$&  Time (s)\\
\hline
{\bf{TenSeg}} &  &  & - & - & - & {\bf{500}} & 0 & 0 & 0 & - & 0.208\\
{\bf{sbs}} &  &  & - & - & - & {\bf{497}} & 3 & 0 & 0 & - & 5.484\\
{\bf{BO}} &  &  & - & - & - & {\bf{497}} & 3 & 0 & 0 & - & 5.193\\
Ratio & CP0 & 20 & - & - & - & 450 & 46 & 4 & 0 & - & 0.067\\
WYR &  &  & - & - & - & 15 & 29 & 86 & 370 & - & 0.132\\
{\bf{GP}} &  &  & - & - & - & {\bf{500}} & 0 & 0 & 0 & - & 0.010\\
\hline
{\bf{TenSeg}} &  &  & - &  - & - & {\bf{499}} & 1 & 0 & 0 & - & 0.055\\
{\bf{sbs}} &  &  & - & - & - & {\bf{500}} & 0 & 0 & 0 & - & 2.028\\
{\bf{BO}} &  &  & - & - & - & {\bf{495}} & 5 & 0 & 0 & - & 3.846\\
Ratio & CP0 & 10 & - &  - & - & 446 & 51 & 3 & 0 & - & 0.065\\
WYR &  &  & - & - & - & 2 & 2 & 7 & 489 & - & 0.155\\
{\bf{GP}} &  &  & - & - & - & {\bf{500}} & 0 & 0 & 0 & - & 0.007\\
\hline
{\bf{TenSeg}} &  &  & - & - & - & {\bf{498}} & 1 & 1 & 0 & - & 0.015\\
{\bf{sbs}} &  &  & - & - & - & {\bf{500}} & 0 & 0 & 0 & - & 1.689\\
{\bf{BO}} &  &  & - & - & - & {\bf{492}} & 8 & 0 & 0 & - & 9.310\\
Ratio & CP0 & 5 & - & - & - & 440 & 54 & 6 & 0 & - & 0.055\\
WYR &  &  & - & - & - & 0 & 0 & 0 & 500 & - & 0.163\\
{\bf{GP}} &  &  & - & - & - & {\bf{500}} & 0 & 0 & 0 & - & 0.005\\
\hline
\hline
{\bf{TenSeg}} &  &  & - & - & 0 & {\bf{496}} & 3 & 0 & 1 & 0.006 & 0.241\\
{\bf{sbs}} &  &  & - & - & 0 & {\bf{498}} & 2 & 0 & 0 & 0.005 & 5.912\\
BO &  &  & - & - & 0 & 412 & 86 & 2 & 0 & 0.014 & 10.230\\
Ratio & CP1 & 20 & - & - & 0 & 390 & 98 & 12 & 0 & 0.052 & 0.096\\
WYR &  &  & - & - & 0 & 41 & 291 & 167 & 1 & 0.259 & 0.095\\
{\bf{GP}} &  &  & - & - & 0 & {\bf{500}} & 0 & 0 & 0 & 0.001 & 0.015\\
\hline
{\bf{TenSeg}} &  &  & - &  - & 0 & {\bf{498}} & 2 & 0 & 0 & 0.007 & 0.047\\
{\bf{sbs}} &  &  & - & - & 0 & {\bf{500}} & 0 & 0 & 0 & 0.004 & 2.305\\
BO &  &  & - & - & 0 & 416& 77 & 7 & 0 & 0.019 & 7.663\\
Ratio & CP1 & 10 & - &  - & 0 & 383 & 98 & 17 & 2 & 0.055 & 0.105\\
WYR &  &  & - & - & 0 & 3 & 10 & 161 & 326 & 0.359 & 0.114\\
{\bf{GP}} &  &  & - & - & 0 & {\bf{500}} & 0 & 0 & 0 & 0.001 & 0.013\\
\hline
{\bf{TenSeg}} &  &  & - & - & 0 & {\bf{498}} & 2 & 0 & 0 & 0.006 & 0.013\\
{\bf{sbs}} &  &  & - & - & 0 & {\bf{500}} & 0 & 0 & 0 & 0.004 & 2.039\\
BO &  &  & - & - & 0 & 393 & 102 & 5 & 0 & 0.023 & 9.310\\
Ratio & CP1 & 5 & - & - & 0 & 367 & 107 & 22 & 4 & 0.072 & 0.121\\
WYR &  &  & - & - & 0 & 0 & 1 & 9 & 490 & 0.395 & 0.144\\
{\bf{GP}} &  &  & - & - & 0 & {\bf{489}} & 11 & 0 & 0 & 0.002& 0.011\\
\hline
\end{tabular}}}
\label{table:results_AR_app}
\end{table}
\begin{table}
\centering
\caption{Simulation results for the ER model with 0 (CP0) and 1 (CP1) true change-points.  We used 20, 10, 5 number of components in the CP decomposition. We provide the distribution of $\hat{N} - N$ over 500 simulation iterations, the average Hausdorff distance, $d_H$, and the average computational time (s) for each method.}
\vspace{0.05in}
{\small{
\begin{tabular}{|l|l|l|l|l|l|l|l|l|l|l|l|}
\cline{1-12}
\vspace{-0.1in}
 &  &  & \multicolumn{7}{|c|}{} &  & \\
 &  &  & \multicolumn{7}{|c|}{$\hat{N} - N$} &  & \\
Method & Model & $C_{{\rm CP}}$ & $\leq -3$ & -2 & -1 &0 & 1 & 2 & $\geq 3$ & $d_H$&  Time (s)\\
\hline
{\bf{TenSeg}} &  &  & - & - & - & {\bf{500}} & 0 & 0 & 0 & - & 0.223\\
{\bf{sbs}} &  &  & - & - & - &  {\bf{493}} & 7 & 0 & 0 & - & 3.182\\
{\bf{BO}} &  &  & - & - & - & {\bf{487}} & 13 & 0 & 0 & - & 6.013\\
{\bf{Ratio}} & CP0 & 20 & - & - & - & {\bf{454}} & 42 & 4 & 0 & - & 0.054\\
WYR &  &  & - & - & - & 443 & 55 & 2 & 0 & - & 0.052\\
{\textbf{GP}} &  &  & - & - & - & {\textbf{500}} & 0 & 0 & 0 & - & 0.006\\
\hline
{\bf{TenSeg}} &  &  & - &  - & - & {\bf{500}} & 0 & 0 & 0 & - & 0.051\\
{\bf{sbs}} &  &  & - & - & - & {\bf{500}} & 0 & 0 & 0 & - & 1.482\\
{\bf{BO}} &  &  & - & - & - & {\bf{478}} & 22 & 0 & 0 & - & 4.192\\
{\bf{Ratio}} & CP0 & 10 & - &  - & - & {\bf{451}} & 43 & 6 & 0 & - & 0.058\\
WYR &  &  & - & - & - & 184 & 124 & 92 & 100 & - & 0.089\\
{\textbf{GP}} &  &  & - & - & - & {\textbf{500}} & 0 & 0 & 0 & - & 0.004\\
\hline
{\bf{TenSeg}} &  &  & - & - & - & {\bf{495}} & 5 & 0 & 0 & - & 0.016\\
{\bf{sbs}} &  &  & - & - & - & {\bf{500}} & 0 & 0 & 0 & - & 1.137\\
{\bf{BO}} &  &  & - & - & - & {\bf{462}} & 38 & 0 & 0 & - & 4.002\\
Ratio & CP0 & 5 & - & - & - & 432 & 56 & 12 & 0 & - & 0.061\\
WYR &  &  & - & - & - & 28 & 28 & 43 & 401 & - & 0.148\\
{\textbf{GP}} &  &  & - & - & - & {\textbf{500}} & 0 & 0 & 0 & - & 0.004\\
\hline
\hline
{\bf{TenSeg}} &  &  & - & - & 0 & {\bf{479}} & 15 & 5 & 1 & 0.003 & 0.225\\
{\bf{sbs}} &  &  & - & - & 0 & {\bf{498}} & 2 & 0 & 0 & 0.001 & 3.927\\
{\bf{BO}} &  &  & - & - & 0 & {\bf{453}} & 44 & 3 & 0 & 0.006 & 12.024\\
Ratio & CP1 & 20 & - & - & 0 & 380 & 108 & 10 & 2 & 0.057 & 0.080\\
WYR &  &  & - & - & 0 & 9 & 269 & 222 & 0 & 0.278 & 0.081\\
{\bf{GP}} &  &  & - & - & 0 & {\bf{500}} & 0 & 0 & 0 & 0.001 & 0.012\\
\hline
{\bf{TenSeg}} &  &  & - &  - & 0 & {\bf{487}} & 13 & 0 & 0 & 0.001 & 0.055\\
{\bf{sbs}} &  &  & - & - & 0 & {\bf{497}} & 3 & 0 & 0 & 0.002 & 1.702\\
BO &  &  & - & - & 0 & 418 & 77 & 5 & 0 & 0.023 & 7.811\\
Ratio & CP1 & 10 & - &  - & 0 & 397 & 85 & 17 & 1 & 0.053 & 0.098\\
WYR &  &  & - & - & 0 & 1 & 3 & 142 & 354 & 0.358 & 0.110\\
{\bf{GP}} &  &  & - & - & 0 & {\bf{500}} & 0 & 0 & 0 & 0.001 & 0.009\\
\hline
{\bf{TenSeg}} &  &  & - & - & 0 & {\bf{492}} & 8 & 0 & 0 & 0.003 & 0.018\\
{\bf{sbs}} &  &  & - & - & 0 & {\bf{496}} & 4 & 0 & 0 & 0.003 & 1.194\\
BO &  &  & - & - & 0 & 418 & 71 & 11 & 0 & 0.023 & 7.812\\
Ratio & CP1 & 5 & - & - & 0 & 379 & 102 & 16 & 3 & 0.056 & 0.103\\
WYR &  &  & - & - & 0 & 0 & 0 & 2 & 498 & 0.395 & 0.122\\
{\bf{GP}} &  &  & - & - & 0 & {\bf{500}} & 0 & 0 & 0 & 0.002& 0.008\\
\hline
\end{tabular}}}
\label{table:results_ER_app}
\end{table}
\begin{table}
\centering
\caption{Simulation results for the SB model with 0 (CP0) and 1 (CP1) true change-points.  We used 20, 10, 5 number of components in the CP decomposition. We provide the distribution of $\hat{N} - N$ over 500 simulation iterations, the average Hausdorff distance, $d_H$, and the average computational time (s) for each method.}
\vspace{0.05in}
{\small{
\begin{tabular}{|l|l|l|l|l|l|l|l|l|l|l|l|}
\cline{1-12}
 &  &  & \multicolumn{7}{|c|}{} &  & \\
 &  &  & \multicolumn{7}{|c|}{$\hat{N} - N$} &  & \\
Method & Model & $C_{{\rm CP}}$ & $\leq -3$ & -2 & -1 &0 & 1 & 2 & $\geq 3$ & $d_H$&  Time (s)\\
\hline
\textbf{TenSeg} &  &  & - & - & - & \textbf{500} & 0 & 0 & 0 & - & 0.242\\
\textbf{sbs} &  &  & - & - & - & \textbf{500} & 0 & 0 & 0 & - & 3.249\\
\textbf{BO}&  &  & - & - & - & \textbf{500} & 0 & 0 & 0 & - & 4.691\\
{\bf{Ratio}} & CP0 & 20 & - & - & - & {\bf{451}} & 44 & 4 & 1 & - & 0.065\\
WYR &  &  & - & - & - & 407 & 75 & 16 & 2 & - & 0.069\\
{\bf{GP}} &  &  & - & - & - & {\bf{500}} & 0 & 0 & 0 & - & 0.009\\
\hline
\textbf{TenSeg} &  &  & - &  - & - & \textbf{500} & 0 & 0 & 0 & - & 0.062\\
\textbf{sbs} &  &  & - & - & - & \textbf{500} & 0 & 0 & 0 & - & 1.701\\
\textbf{BO} &  &  & - & - & - & \textbf{494} & 6 & 0 & 0 & - & 4.124\\
{\textbf{Ratio}} & CP0 & 10 & - &  - & - & {\textbf{451}} & 45 & 4 & 0 & - & 0.056\\
WYR &  &  & - & - & - & 84 & 76 & 102 & 238 & - & 0.113\\
{\bf{GP}} &  &  & - & - & - & {\bf{500}} & 0 & 0 & 0 & - & 0.004\\
\hline
\textbf{TenSeg} &  &  & - & - & - & \textbf{496} & 4 & 0 & 0 & - & 0.019\\
\textbf{sbs} &  &  & - & - & - & \textbf{500} & 0 & 0 & 0 & - & 1.319\\
\textbf{BO} &  &  & - & - & - & \textbf{487} & 13 & 0 & 0 & - & 3.972\\
Ratio & CP0 & 5 & - & - & - & 428 & 63 & 7 & 2 & - & 0.059\\
WYR &  &  & - & - & - & 7 & 12 & 13 & 468 & - & 0.162\\
{\bf{GP}} &  &  & - & - & - & {\bf{500}} & 0 & 0 & 0 & - & 0.004\\
\hline
\hline
\textbf{TenSeg} &  &  & - & - & 0 & \textbf{467} & 29 & 2 & 2 & 0.004 & 0.241\\
\textbf{sbs} &  &  & - & - & 0 & \textbf{500} & 0 & 0 & 0 & 0.001 & 3.301\\
BO &  &  & - & - & 0 & 437 & 56 & 7 & 0 & 0.011 & 8.995\\
Ratio & CP1 & 20 & - & - & 0 & 398 & 98 & 4 & 0 & 0.045 & 0.097\\
WYR &  &  & - & - & 0 & 8 & 248 & 243 & 1 & 0.283 & 0.098\\
{\bf{GP}} &  &  & - & - & 0 & {\bf{500}} & 0 & 0 & 0 & 0.001 & 0.015\\
\hline
\textbf{TenSeg} &  &  & - &  - & 0 & \textbf{493} & 6 & 1 & 0 & 0.001 & 0.055\\
\textbf{sbs} &  &  & - & - & 0 & \textbf{500}& 0 & 0 & 0 & 0.001 & 1.723\\
BO &  &  & - & - & 0 & 416 & 79 & 5 & 0 & 0.019 & 7.724\\
Ratio & CP1 & 10 & - &  - & 0 & 403 & 85 & 12 & 0 & 0.046 & 0.095\\
WYR &  &  & - & - & 0 & 0 & 4 & 95 & 401 & 0.363 & 0.108\\
{\bf{GP}} &  &  & - & - & 0 & {\bf{500}} & 0 & 0 & 0 & 0.001 & 0.009\\
\hline
\textbf{TenSeg} &  &  & - & - & 0 & \textbf{492} & 8 & 0 & 0 & 0.003 & 0.019\\
\textbf{sbs} &  &  & - & - & 0 & \textbf{500} & 0 & 0 & 0 & 0.003 & 1.314\\
BO &  &  & - & - & 0 & 428 & 71 & 1 & 0 & 0.012 & 7.410\\
Ratio & CP1 & 5 & - & - & 0 & 379 & 98 & 21 & 2 & 0.060 & 0.103\\
WYR &  &  & - & - & 0 & 0 & 0 & 1 & 499 & 0.396 & 0.125\\
{\bf{GP}} &  &  & - & - & 0 & {\bf{495}} & 5 & 0 & 0 & 0.002& 0.008\\
\hline
\end{tabular}}}
\label{table:results_SB_app}
\end{table}
\begin{table}
\centering
\caption{Simulation results for the AR model with smaller change magnitudes (equal to 0.2, instead of 0.6 used in Table \ref{table:results_AR_app}) when we have 1 true change-points.  We used 20, 10, 5 number of components in the CP decomposition. We provide the distribution of $\hat{N} - N$ over 500 simulation iterations, the average Hausdorff distance, $d_H$, and the average computational time (s) for each method.}
\vspace{0.05in}
{\small{
\begin{tabular}{|l|l|l|l|l|l|l|l|l|l|l|l|}
\cline{1-12}
 &  &  & \multicolumn{7}{|c|}{} &  & \\
 &  &  & \multicolumn{7}{|c|}{$\hat{N} - N$} &  & \\
Method & Model & $C_{{\rm CP}}$ & $\leq -3$ & -2 & -1 &0 & 1 & 2 & $\geq 3$ & $d_H$&  Time (s)\\
\hline
\textbf{TenSeg} &  &  & - & - & 0 & \textbf{479} & 11 & 10 & 0 & 0.014 & 0.213\\
\textbf{sbs} &  &  & - & - & 0 & \textbf{500} & 0 & 0 & 0 & 0.003 & 5.421\\
BO &  &  & - & - & 0 & 431 & 56 & 13 & 0 & 0.013 & 6.619\\
Ratio & CP1 & 20 & - & - & 0 & 380 & 107 & 13 & 0 & 0.060 & 0.087\\
WYR &  &  & - & - & 0 & 59 & 329 & 112 & 0 & 0.235 & 0.090\\
{\bf{GP}} &  &  & - & - & 0 & {\bf{500}} & 0 & 0 & 0 & 0.001 & 0.012\\
\hline
\textbf{TenSeg} &  &  & - &  - & 0 & \textbf{495} & 5 & 0 & 0 & 0.008 & 0.039\\
\textbf{sbs} &  &  & - & - & 0 & \textbf{500}& 0 & 0 & 0 & 0.005 & 2.512\\
BO &  &  & - & - & 0 & 421 & 70 & 9 & 0 & 0.020 & 4.329\\
Ratio & CP1 & 10 & - &  - & 0 & 377 & 113 & 10 & 0 & 0.057 & 0.088\\
WYR &  &  & - & - & 0 & 0 & 19 & 165 & 316 & 0.360 & 0.100\\
{\bf{GP}} &  &  & - & - & 0 & {\bf{500}} & 0 & 0 & 0 & 0.002 & 0.011\\
\hline
\textbf{TenSeg} &  &  & - & - & 0 & \textbf{498} & 2 & 0 & 0 & 0.013 & 0.010\\
\textbf{sbs} &  &  & - & - & 0 & \textbf{500} & 0 & 0 & 0 & 0.006 & 1.764\\
BO &  &  & - & - & 0 & 420 & 76 & 4 & 0 & 0.022 & 4.212\\
Ratio & CP1 & 5 & - & - & 0 & 379 & 97 & 24 & 0 & 0.061 & 0.109\\
WYR &  &  & - & - & 0 & 0 & 0 & 2 & 498 & 0.398 & 0.133\\
{\bf{GP}} &  &  & - & - & 0 & {\bf{498}} & 2 & 0 & 0 & 0.004& 0.011\\
\hline
\end{tabular}}}
\label{table:results_AR_app_diff}
\end{table}}
\section{Enron data}
\label{sec:res_enron}
We now discuss the results from applying TenSeg to  the Carnegie Mellon University CALO Project Enron email dataset, which is extensively discussed in Section 5.2 of the main article.  The NORMO algorithm was first employed in order to select the number of components in the CP decomposition. In addition to the estimated locations of the change-points obtained by TenSeg, we also provide a possible explanation of the results. TenSeg detected thirteen change-points that correspond to the weeks starting on the
\begin{itemize}
\item May 10, 1999: This change-point is most certainly related to the so-called Silverpeak Incident, which occurred just two time-points later. More precisely, on May 24, Tim Belden, head of Enron's West Coast Trading Desk in Portland, Oregon, conducts his first experiment to exploit the new rules of California's deregulated energy market. He creates congestion on power lines which causes electricity prices to rise to \$7 million for California. 

\item September 20, 1999: During the previous week, and more specifically, on September 16, 1999, Enron's CFO Andy Fastow addresses Merrill Lynch in the World Trade Center, asking the team of investment bankers to find investors for his LJM2 Fund. He assures them: ``If there's a conflict between Enron and LJM, I will favor LJM''.

\item November 29, 1999: This change-point could be associated with two important events that took place in mid-October, which is six time-points prior to the detected change-point. On October 12, the Enron board exempts the CFO from Enron's ``code of ethics'' so that he can raise money for LJM2, and the next day, Merrill Lynch releases a placement memo for LJM2.

\item February 7, 2000: Two weeks, meaning two time-points, before this detection, the Annual Analysts Meeting took place. After important statements made by Jeff Skilling, Enron's COO, by the end of the day, Enron's stock rose 26\% to a new high of \$67.25.

\item May 29, 2000: This detection is related to an event that occurred one week earlier. The California ISO (Independent System Operator), the organization in charge of California's electricity supply and demand, declared a Stage One Emergency, warning of low power reserves.

\item July 3, 2000: In July 2000, Enron announced that its Broadband unit (EBS) joined forces with Blockbuster to supply video-on-demand. 

\item August 21, 2000: This detection could be related to the fact that on August 23, Enron's stock hit an all-time high of \$90.56 with a market valuation of \$70 billion. Furthermore, the Federal Energy Regulatory Commission ordered an investigation into strategies designed to drive electricity prices up in California.

\item October 30, 2000: This detection is most probably related to the fact that on November 1, 2000, the Federal Energy Regulatory Commission (FERC) investigation exonerated Enron for any wrongdoing in the state of California.

\item April 30, 2001: This change-point could be related to two events that took place in its vicinity. Firstly, at the end of March, 2001, Enron schedules an unusual analyst conference call in order to boost stock, which actually works. Secondly in mid April, 2001, a quite eventful Quarterly Conference Call took place, which plays a big part in Enron’s stock falling from around \$90 per share eight months prior to just \$60 at the time after the call. 

\item July 16, 2001: This detection is related to an event that takes place on July 13. The CEO, Jeff Skilling, announces his desire to resign to the chairman, Kenneth Lay.

\item September 17, 2001: During this month, Enron's former CEO sells \$15.5 million of stock, bringing stock sales since May 2000 to over \$70 million.

\item December 3, 2001: November 2001 was a month full of major events relating to the company's future. Firstly, on November 8, Enron starts negotiations to sell itself to Dynegy, a smaller rival, in order to head off bankruptcy; the next day, Dynegy agrees to buy Enron for about \$9 billion in stock and cash. On November 19, 2001, the company restates its third quarter earnings and discloses it is trying to restructure a \$690 million obligation, while on November 28, the company's shares plunge below \$1; as a result, Dynegy withdraws from the deal the next day. On December 2, Enron files for Chapter 11 bankruptcy protection, at the time the largest bankruptcy in US history.

\item February 11, 2002: From late January until mid-February 2002, a series of serious events take place. First, the Justice Department confirms it has begun a criminal investigation into Enron. Second, the founder and chairman of the company resigns, and the former Enron vice chairman commits suicide. Third, an internal investigation into Enron's collapse led by the University of Texas School of Law Dean spreads blame among self-dealing executives and negligent directors. Fourth, in February, many of the former CEOs testify before a Congressional panel; the founder of the company refuses to do so.
\end{itemize}
As previously stated, the Enron email dataset has been widely analyzed in the literature, but this is the first time that the 33 different topic categorizations have been (and could have been) considered jointly as a tensor object. Previously, in a series of articles (e.g., \citealp{Priebe2006, Park2012, Peel2015, DeRidder2016, Luo2023}), the object examined for change-point detection was of dimensionality $184 \times 184$ at each one of the 189 time-points but never using information across all topics. In addition, \cite{huang2022} treated the e-mail exchange data set on one topic as a 2-way tensor and considered changes in its mean structure.  They detected change-points in July 1999, August 2000 and July 2001 using their SFD statistic and in November 1999, December 2000 and September 2001 using their MSFD statistic. By merging change-points (located across the different article references above) that are within a distance of at most 4 time points (nearly a month), the following 11 change-point locations are in at least one of the references above: 05/24/1999, 06/28/1999, 09/20/1999, 12/13/1999, 05/15/2000, 08/21/2000, 12/11/2000, 08/13/2001, 10/22/2001, 12/03/2001, 03/11/2002. The two most prominent ones, indicated in the majority of those articles are those referring to the weeks starting on 05/15/2000 and 08/21/2000.  

We conclude that treating the object as a tensor allows for added detection capability.  In particular, by considering all the 33 different topics in the email exchange networks we are still able to capture the important change-points or movements, but we can additionally shed some more light. For example, none of the previous works (where the data were treated as directed graphs instead of tensors) managed to detect some important change-points, such as of the week starting on 02/07/2000 (Enron's stock rose by 26\%), on 07/03/2000 (EBS joined forces with Blockbuster), or on 04/30/2001 (huge fall in Enron's stock price).  This shows the power of the TenSeg method and its generalizability.

\section{Tables discussed in Sections 6.1 and 6.2 of the main article}
\label{sec:res_auto_HOSVD}
In this section, we provide the tables with the results discussed in Sections 6.1 and 6.2 for the performance of our method on autocorrelated data (Table \ref{table:results_AR_correlated}) and when the HOSVD (instead of CP) decomposition method was employed (Table \ref{table:results_AR_hosvd}), respectively. The relevant \textsf{R} code to reproduce the results of this section can be found at \url{https://github.com/Anastasiou-Andreas/TenSeg}.

{\setstretch{1}
\begin{table}
\centering
\caption{Simulation results for the AR, ER, and SB types of changes with 0 (CP0), 1 (CP1), 4 (CP4) and 10 (CP10) true change-points together with serially correlated data; the correlation coefficient of the AR(1) process for the noise is equal to $\alpha = 0.7$. We used 20, 10, 5 number of components in the CP decomposition. We provide the distribution of $\hat{N} - N$ over 500 simulation iterations, the average Hausdorff distance, $d_H$, and the average computational time (s) for each method.}
\vspace{0.05in}
{\small{
\begin{tabular}{|c|l|l|l|l|l|l|l|l|l|l|l|}
\cline{1-12}
 & &  &  \multicolumn{7}{|c|}{} &  & \\
 & &  &  \multicolumn{7}{|c|}{$\hat{N} - N$} &  & \\
Precision matrix & Model & $C_{{\rm CP}}$ & $\leq -3$ & -2 & -1 &0 & 1 & 2 & $\geq 3$ & $d_H$&  Time (s)\\
\hline
 & & 20 & - & - & - & 500 & 0 & 0 & 0 & - & 0.143\\
 & CP0 & 10 & - &  - & - & 500 & 0 & 0 & 0 & - & 0.051\\
 & & 5 & - & - & - & 500 & 0 & 0 & 0 & - & 0.016\\
\cline{2-12}
 & & 20 & - & - & 0 & 499 & 1 & 0 & 0 & 0.005 & 0.109\\
 & CP1 & 10 & - &  - & 0 & 500 & 0 & 0 & 0 & 0.005 & 0.039\\
AR & & 5 & - & - & 0 & 500 & 0 & 0 & 0 & 0.005 & 0.014\\
\cline{2-12}
 & & 20 & 0 & 0 & 0 & 498 & 0 & 2 & 0 & 0.003 & 0.602\\
 & CP4 & 10 & 0 & 0 & 0 & 498 & 1 & 1 & 0 & 0.003 & 0.125\\
 & & 5 & 0 & 0 & 0 & 487 & 13 & 0 & 0 & 0.003 & 0.032\\
\cline{2-12}
 & & 20 & 0 & 0 & 0 & 500 & 0 & 0 & 0 & 0.002 & 1.510\\
 & CP10 & 10 & 0& 0 &0  & 499 & 1 & 0 & 0 & 0.002 & 0.322\\
 & & 5 & 0 & 0 & 0 & 492 & 8 & 0 & 0 & 0.002 & 0.074\\
\hline
\hline
 & & 20 & - & - & - & 500 & 0 & 0 & 0 & - & 0.158\\
 & CP0 & 10 & - &  - & - & 500 & 0 & 0 & 0 & - & 0.043\\
 & & 5 & - & - & - & 500 & 0 & 0 & 0 & - & 0.014\\
\cline{2-12}
 & & 20 & - & - & 0 & 490 & 7 & 3 & 0 & 0.001 & 0.118\\
 & CP1 & 10 & - &  - & 0 & 498 & 2 & 0 & 0 & 0.001 & 0.032\\
SB & & 5 & - & - & 0 & 496 & 4 & 0 & 0 & 0.001 & 0.011\\
\cline{2-12}
 & & 20 & 0 & 0 & 0 & 499 & 1 & 0 & 0 & 0.003 & 0.636\\
 & CP4 & 10 & 0 & 0 & 0 & 500 & 0 & 0 & 0 & 0.003 & 0.099\\
 & & 5 & 0 & 0 & 0 & 489 & 11 & 0 & 0 & 0.003 & 0.026\\
\cline{2-12}
 & & 20 & 0 & 0 & 0 & 498 & 2 & 0 & 0 & 0.002 & 2.049\\
 & CP10 & 10 & 0& 0 &0 & 500 & 0 & 0 & 0 & 0.001 & 0.322\\
 & & 5 & 0 & 0 & 0 & 492 & 7 & 1 & 0 & 0.002 & 0.066\\
\hline
\hline
 & & 20 & - & - & - & 500 & 0 & 0 & 0 & - & 0.147\\
 & CP0 & 10 & - &  - & - & 500 & 0 & 0 & 0 & - & 0.040\\
 & & 5 & - & - & - & 500 & 0 & 0 & 0 & - & 0.014\\
\cline{2-12}
 & & 20 & - & - & 0 & 499 & 1 & 0 & 0 & 0.001 & 0.115\\
 & CP1 & 10 & - &  - & 0 & 498 & 2 & 0 & 0 & 0.001 & 0.032\\
ER & & 5 & - & - & 0 & 493 & 7 & 0 & 0 & 0.001 & 0.013\\
\cline{2-12}
 & & 20 & 2 & 29 & 81 & 311 & 67 & 10 & 0 & 0.051 & 0.265\\
 & CP4 & 10 & 30 & 85 & 191 & 163 & 31 & 0 & 0 & 0.143 & 0.096\\
 & & 5 & 78 & 142 & 175 & 103 & 2 & 0 & 0 & 0.219 & 0.021\\
\cline{2-12}
 & & 20 & 2 & 14 & 47 & 289 & 122 & 25 & 1 & 0.026 & 0.927\\
 & CP10 & 10 & 42 & 58 & 134 & 230 & 34 & 2 & 0 & 0.069 & 0.221\\
 & & 5 & 274 & 115 & 71 & 37 & 3 & 0 & 0 & 0.139 & 0.056\\
\hline
\end{tabular}}}
\label{table:results_AR_correlated}
\end{table}

{\setstretch{1}
\begin{table}
\centering
\caption{Simulation results for the AR, SB, and ER types of changes with 0 (CP0), 1 (CP1), 4 (CP4) and 10 (CP10) true change-points when, instead of CP, the HOSVD decomposition method has been employed. We used 20, 10, 5 number of components in the HOSVD decomposition.  We provide the distribution of $\hat{N} - N$ over 500 simulation iterations, the average Hausdorff distance, $d_H$, and the average computational time (s) for each method.}
\vspace{0.05in}
{\small{
\begin{tabular}{|c|l|l|l|l|l|l|l|l|l|l|l|}
\cline{1-12}
&  &  & \multicolumn{7}{|c|}{} &  & \\
&  &  & \multicolumn{7}{|c|}{$\hat{N} - N$} &  & \\
Precision matrix & Model & Components & $\leq -3$ & -2 & -1 &0 & 1 & 2 & $\geq 3$ & $d_H$&  Time (s)\\
\hline
& & 20 & - & - & - & 500 & 0 & 0 & 0 & - & 0.234\\
& CP0 & 10 & - & - & - & 500 & 0 & 0 & 0 & - & 0.056\\
& & 5 & - & - & - & 495 & 4 & 1 & 0 & - & 0.016\\
\cline{2-12}
& & 20 & - & - & 0 & 495 & 4 & 1 & 0 & 0.007 & 0.226\\
& CP1 & 10 & - & - & 0 & 497 & 3 & 0 & 0 & 0.006 & 0.048\\
AR & & 5 & - & - & 0 & 497 & 3 & 0 & 0 & 0.007 & 0.013\\
\cline{2-12}
& & 20 & 0 & 0 & 0 & 482 & 10 & 4 & 4 & 0.006 & 1.194\\
& CP4 & 10 & 0 & 0 & 0 & 493 & 4 & 3 & 0 & 0.005 & 0.235\\
& & 5 & 0 & 0 & 0 & 464 & 29 & 7 & 0 & 0.009 & 0.051\\
\cline{2-12}
& & 20 & 0 & 0 & 0 & 493 & 2 & 4 & 1 & 0.002 & 5.011\\
& CP10 & 10 & 0 & 0 & 0 & 496 & 3 & 1 & 0 & 0.002 & 0.825\\
& & 5 & 0 & 0 & 0 & 464 & 31 & 3 & 2 & 0.004 & 0.140\\
\hline
\hline
& & 20 & - & - & - & 500 & 0 & 0 & 0 & - & 0.214\\
& CP0 & 10 & - & - & - & 500 & 0 & 0 & 0 & - & 0.041\\
& & 5 & - & - & - & 496 & 4 & 0 & 0 & - & 0.015\\
\cline{2-12}
& & 20 & - & - & 0 & 472 & 20 & 5 & 3 & 0.007 & 0.428\\
& CP1 & 10 & - & - & 0 & 490 & 9 & 1 & 0 & 0.001 & 0.078\\
SB & & 5 & - & - & 0 & 494 & 6 & 0 & 0 & 0.001 & 0.022\\
\cline{2-12}
& & 20 & 0 & 0 & 0 & 483 & 3 & 7 & 7 & 0.007 & 1.076\\
& CP4 & 10 & 0 & 0 & 0 & 495 & 4 & 0 & 1 & 0.004 & 0.139\\
& & 5 & 0 & 0 & 0 & 470 & 29 & 0 & 1 & 0.008 & 0.028\\
\cline{2-12}
& & 20 & 0 & 0 & 0 & 493 & 1 & 0 & 6 & 0.002 & 3.581\\
& CP10 & 10 & 0 & 0 & 0 & 499 & 0 & 1 & 0 & 0.002 & 0.647\\
& & 5 & 0 & 0 & 0 & 447 & 49 & 4 & 0 & 0.006 & 0.115\\
\hline
\hline
& & 20 & - & - & - & 500 & 0 & 0 & 0 & - & 0.212\\
& CP0 & 10 & - & - & - & 500 & 0 & 0 & 0 & - & 0.058\\
& & 5 & - & - & - & 496 & 4 & 0 & 0 & - & 0.014\\
\cline{2-12}
& & 20 & - & - & 0 & 486 & 12 & 2 & 0 & 0.006 & 0.252\\
& CP1 & 10 & - & - & 0 & 494 & 5 & 1 & 0 & 0.004 & 0.058\\
ER & & 5 & - & - & 0 & 489 & 10 & 1 & 0 & 0.008 & 0.017\\
\cline{2-12}
& & 20 & 3 & 14 & 35 & 411 & 31 & 5 & 1 & 0.052 & 0.728\\
& CP4 & 10 & 9 & 24 & 79 & 374 & 14 & 0 & 0 & 0.083 & 0.133\\
& & 5 & 23 & 81 & 134 & 247 & 15 & 0 & 0 & 0.145 & 0.034\\
\cline{2-12}
& & 20 & 0 & 1 & 9 & 448 & 40 & 1 & 1 & 0.020 & 2.520\\
& CP10 & 10 & 2 & 15 & 28 & 429 & 25 & 1 & 0 & 0.031 & 0.512\\
& & 5 & 26 & 64 & 101 & 265 & 40 & 4 & 0 & 0.066 & 0.095\\
\hline
\end{tabular}}}
\label{table:results_AR_hosvd}
\end{table}}

\end{document}